\documentclass[12pt]{iopart}
\usepackage{epsfig}
\usepackage{graphicx}
\usepackage{graphics}
\usepackage{amssymb}
\usepackage{setstack}
\usepackage{rotate}
\usepackage[english]{babel}
\usepackage{color}

\newcommand{\mfig}[4]{ \psfig{figure=#1,width=#2,height=#3,angle=#4 } }
\begin{document}

\bibliographystyle{unsrt}

\title[Bidimensional fluid system with competing interactions]{Bidimensional fluid system with competing interactions: spontaneous and induced pattern formation}
\author{A. Imperio and L. Reatto}
\address{ Istituto Nazionale di Fisica della Materia and Dipartimento di Fisica, Universit\`a degli Studi di Milano, via Celoria 16, 20133 Milan, Italy}

\begin{abstract}
In this paper we present a study of pattern formation in bidimensional systems with competing short-range attractive and long-range repulsive interactions. The interaction parameters are chosen in such a way to analyse two different situations: the spontaneous pattern formation due to the presence of strong competing interactions on different length scales and the pattern formation as a response to an external modulating potential when the system is close to its Lifshitz point. We compare different Monte Carlo techniques showing that Parallel Tempering technique represents a promising approach to study such systems and we present detailed results for the specific heat and the structural properties. We also present random phase approximation predictions about the spontaneous pattern formation (or microphase separation), as well as linear response theory predictions about the induced pattern formation due to the presence of an external modulating field. In particular we observe that the response of our systems to external fields is much stronger compared to the response of a Lennard-Jones fluid.
\end{abstract}

\section{Introduction}

Spontaneous pattern formation can easily be observed in system far from thermal equilibrium due to energy and matter fluxes (\cite{seul95} and references therein), but it can be observed also in equilibrium states as a result of the competition between interactions operating on different length scales \cite{seul95,sear99,gelbart99,brown01,ghezzi97,piazza03,tanaka02}. Patterns appear both in two and three dimensional systems even if, in the last decades, bidimensional or quasi bidimensional systems have grown in importance also due to the possibility of technological applications. In particular the possibility of manipulating the pattern formation throughout the application of external fields has gained great interest. Typical experiments involve fluid films subject to interfering multiple laser beams or ferrofluids in magnetic fields, so that the external agent can manipulate the morphology of clusters as well as the nature of transitions (e.g. the laser induced freezing) \cite{gotze03,wei98,weber02}.\\
The competing terms are typically due to a short-range attraction, which tends to induce a macroscopic phase separation, and a longer-range repulsion, present in addition to the hard core contribution, which causes a frustration of the system favouring the formation of smaller clusters of particles. The short-range attraction is often explained in terms of van der Waals forces, while the long-range repulsion is often related to effective or actual  dipolar interactions as in Langmuir monolayers and in magnetic films \cite{gelbart99,andelman94,debell00}. Other source of long-range repulsion is a screened Coulomb interaction of charged macroparticles (\cite{ghezzi97,muratov02} and references therein).\\
In this paper we present continuum simulation results of a bidimensional system subject to an effective potential which includes competing interactions. Specific features arise if the repulsion is truly long-range like the dipolar one. We do not explore this aspect since we are interested in generic aspects of competition, so both attraction and repulsion are assumed to have a finite range. At first, the potential parameters have been set to study the spontaneous breaking of symmetry leading to pattern formation (or microphase separation) without external field. Simulations have been done adopting different Monte Carlo techniques; in particular, the Parallel Tempering scheme has been explored. Finally the parameters of interactions have been chosen so to study the pattern formation induced by an external modulating potential, when the fluid is near its Lifshitz point. In fact it is known from previous studies \cite{pini00} that specific features appear in the properties of the system even when the long-range repulsion is not strong enough to give rise to microphase separation and a normal liquid-vapor transition is present. Close to this Lifshitz point the coexistence curve is extremely flat and the region of large compressibility increases enormously compared to the normal case. Under these condition we might expect a very large response to an external modulating field. Both  cases have been analysed in different thermodynamic states.\\
The paper is organized as follows. In section \ref{Microphases: model system and computational details} we discuss and compare different Monte Carlo techniques to simulate systems with competing interactions. Computational results are described in section \ref{Microphases: simulation results}; in section \ref{Microphase: mean field models and phase behavior} we compare simulation data with theoretical predictions based on the random phase approximation and describe a very simple model to explain why we observe different patterns as density increases. Section \ref{Simulations with external modulating potential} is dedicated to the analysis of the pattern induced by an external modulating potential. A comparison with a standard Lennard-Jones fluid response is also shown. In section \ref{conclusions} we draw conclusions.

\section{Microphases: model system and computational details \label{Microphases: model system and computational details}}

The model system studied in this paper is similar to that in \cite{sear99}. Unlike the authors \cite{sear99} we have focused our attention on the thermodynamic and structural features of the system. We have analysed such quantities on a large range of temperatures, obtaining information on the microphase region and also on the transition among the ordered states and the homogeneous one. The particles interact via a spherically symmetric pairwise-additive potential which is a sum of two parts:
\begin{equation}
U(r)= U_{sr}(r)+U_{lr}(r).
\label{kac}
\end{equation}
The short-range contribution $U_{sr}$ is the hard disk potential, while the long-range one is:
\begin{equation}
U_{lr}(r)=-\frac{\epsilon_a\sigma^2}{R_a^2}\exp(-\frac{r}{R_a})
       +\frac{\epsilon_r\sigma^2}{R_r^2}\exp(-\frac{r}{R_r}),\label{Ulong}
\end{equation}
in which $r$ is the interparticle distance, $\sigma$ is the hard disk diameter; the subscript letters $a$ and $r$ refer to the ``attraction'' and ``repulsion'', so  $R_a, R_r$ and $\epsilon_a,\epsilon_r$ are respectively the ranges and the strengths of the long-range interactions. In the first part of the paper we study the case $\epsilon_a=\epsilon_r$, so that $\int_0^{\infty}U_{lr}(r)\,d\protect{\bf{r}}=0$. We have performed few computations of control for the ranges $R_a=2\sigma$ and $R_r=4\sigma$ as in \cite{sear99} but all the results reported here have the interaction ranges equal to $R_a=1 \sigma$ and $R_r= 2\sigma$, in order to reduce the computational cost of simulations without affecting the basic physics of the system.
\begin{figure}[ht!]
\begin{center}
\mfig{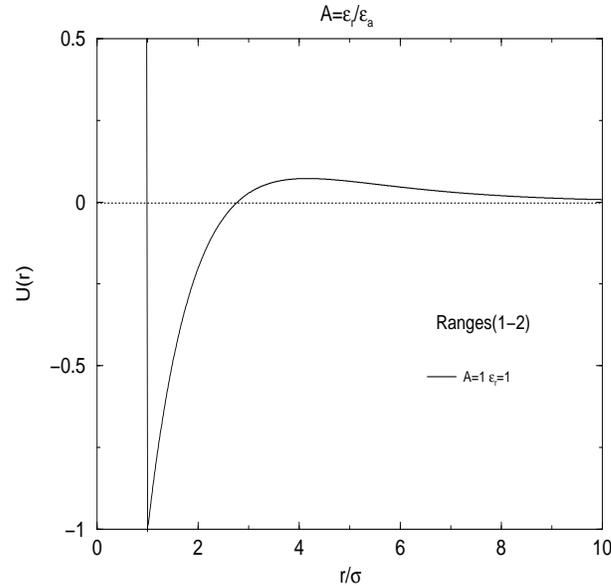}{8cm}{8cm}{0}
\end{center}
\vspace{-1cm}
\caption{\footnotesize{Interaction potential for a pair of particles. The ranges  of the interactions are defined into the text.}}\label{Un_Uo}
\end{figure}
\begin{figure}[h!]
\begin{center}
\mfig{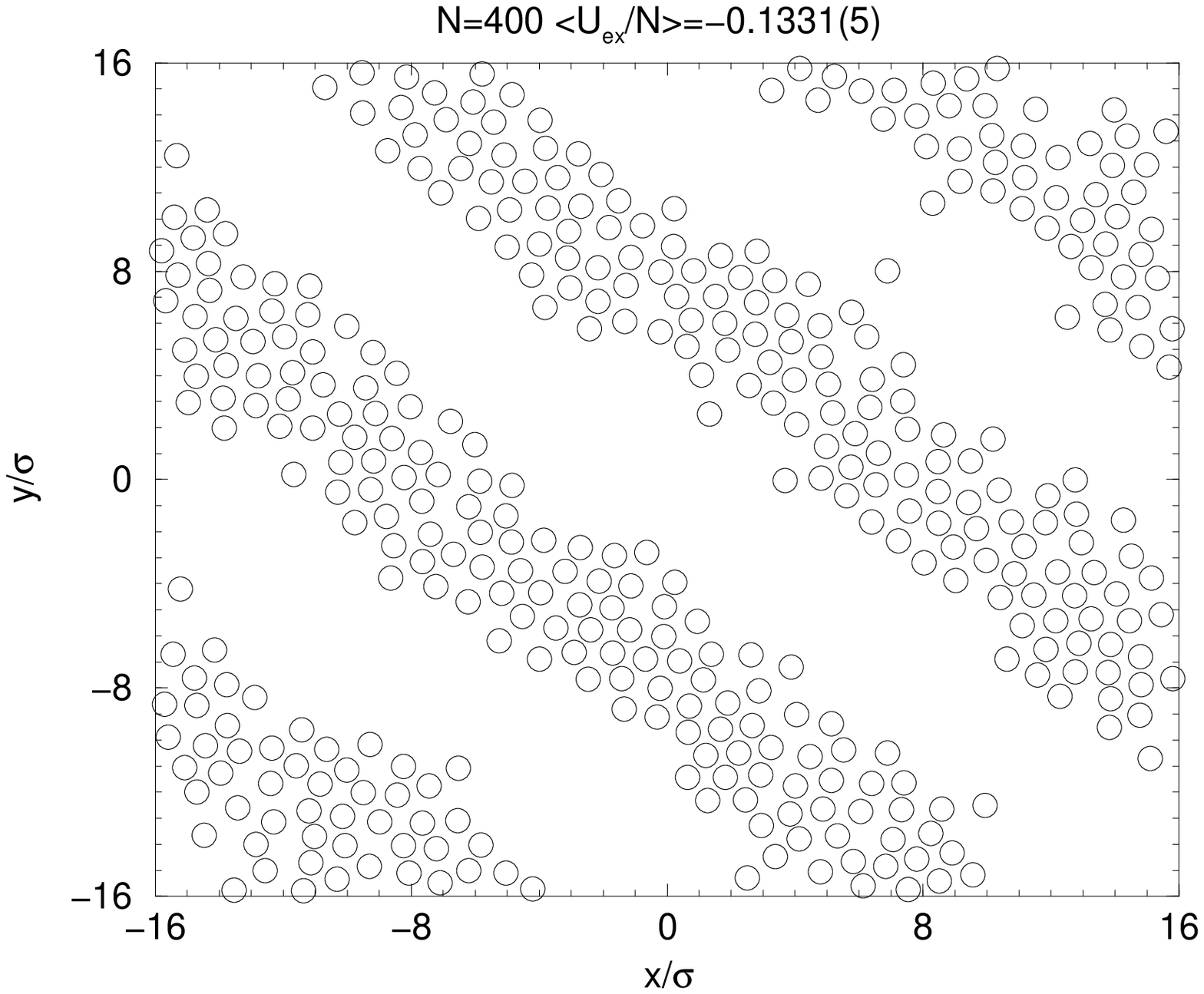}{8cm}{8cm}{0}\mfig{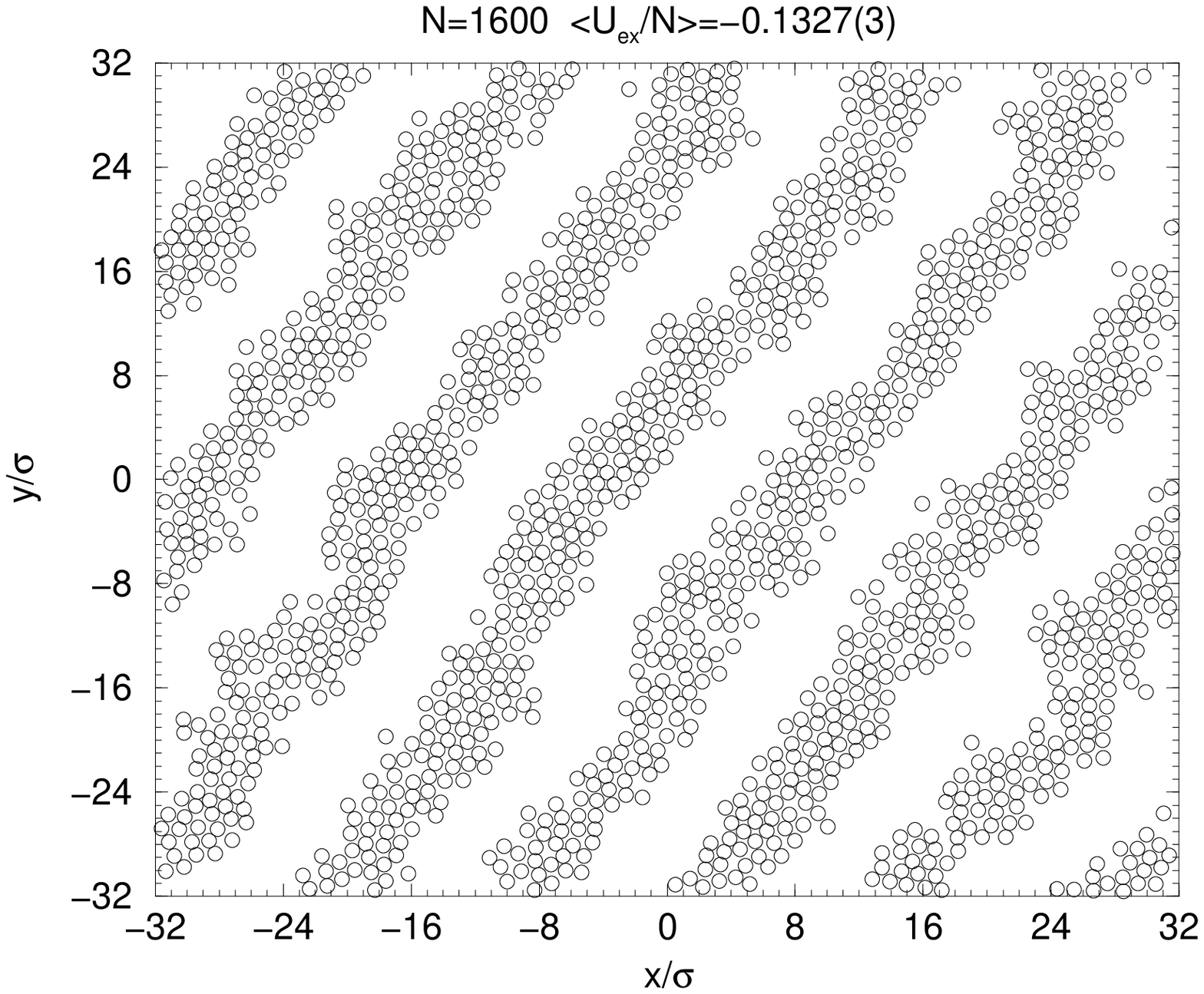}{8cm}{8cm}{0}
\end{center}
\vspace{-1cm}
\caption{\footnotesize{Snapshots relative to the state $\rho=0.4$ $T=0.5$, using different number of particles in MC simulations. There are no relevant modifications to the shape of the striped pattern. The wave vector connected to periodicity of the pattern is the same in both cases: $k_p\sigma=0.60$}}
\label{ss_N}
\end{figure}
In the figures and in all that follows we will use the reduced variables $T\equiv kT/U_c$, $U\equiv U/U_c$ (with $U_c=|U_{lr}(r=\sigma)|$ also referred to as contact value of the potential), $\rho\equiv<\rho>\sigma^2$ (with $<\rho>=N/A$, $N$ number of particles and $A$ simulation box area).\\
We have used different strategies of simulation. 
At first we performed computer simulations using the standard Metropolis Monte Carlo algorithm (in the following MC) \cite{allentildesley} at a fixed number of particles $N$, area $A$ and temperature $T$, with periodic boundary conditions and minimum image convention. We applied a spherical cutoff $R_c$ (always smaller than half of the simulation box side), that is we set the pair potential $U(r)$ to zero for $r\geq R_c$ and then we included tail corrections into the computation of potential energy to compensate for the missing long-range part of $U_{lr}$ due to the cutoff. 
We used $N=400$ but in  preliminary runs we adopted $N=1600$ to test the effect of the finite size of simulation box and in figure \ref{ss_N}, for instance, we have plotted the snapshots relative to a state exhibiting a striped pattern. We can see that the shape of the pattern does not change: stripes remain parallel to each other; the excess internal energy per particle is the same (within the statistical uncertainty); the wave vector which identify the periodicity of the pattern (that we will define more precisely in section \ref{Microphases: simulation results}) is the same in both cases and it corresponds to $k_p\sigma\simeq 0.60$; changes in the radial distribution function computed with different number of particles are less than $2\%$.

\begin{figure}[ht!]
\begin{center}
\mfig{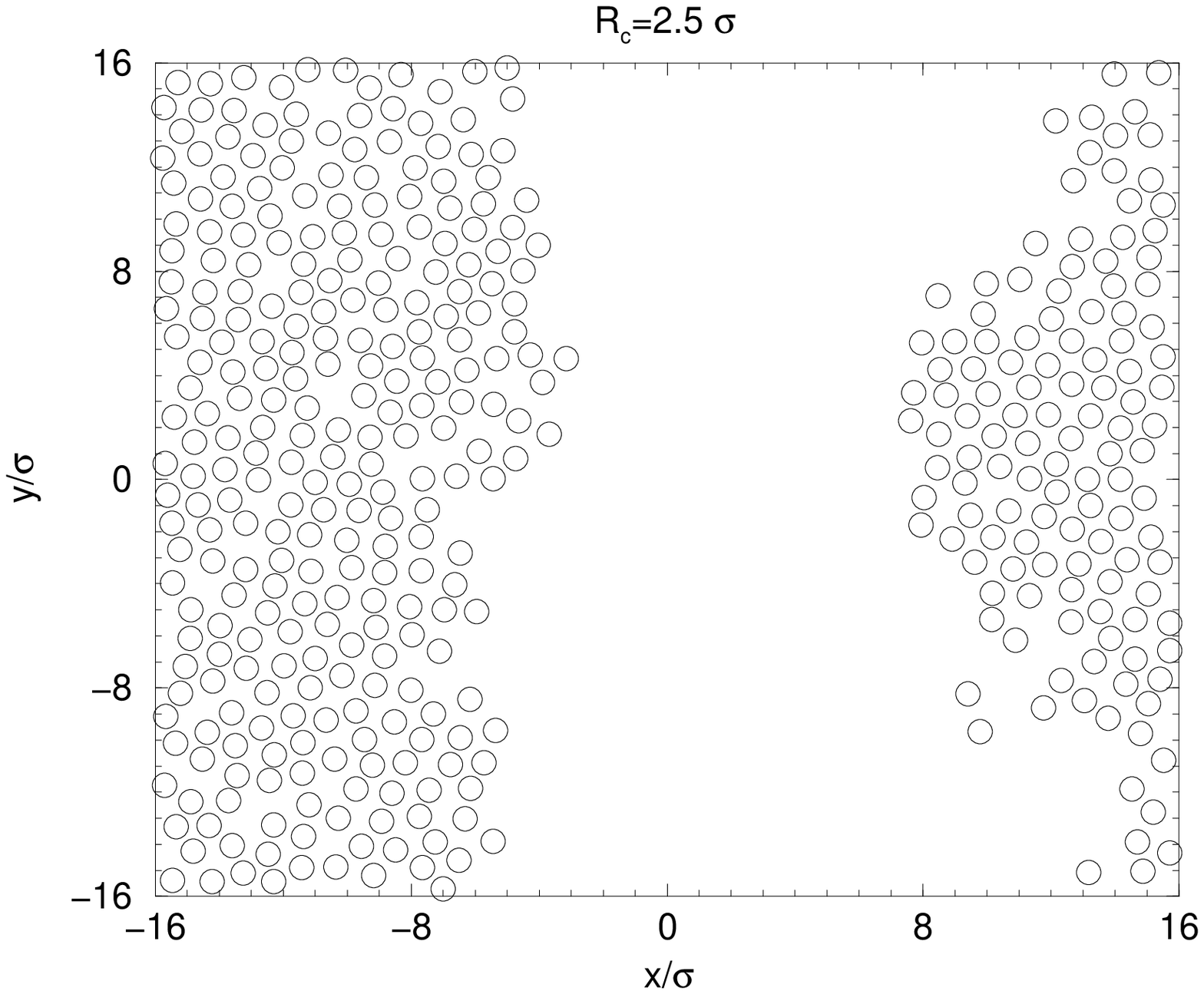}{8cm}{8cm}{0}\mfig{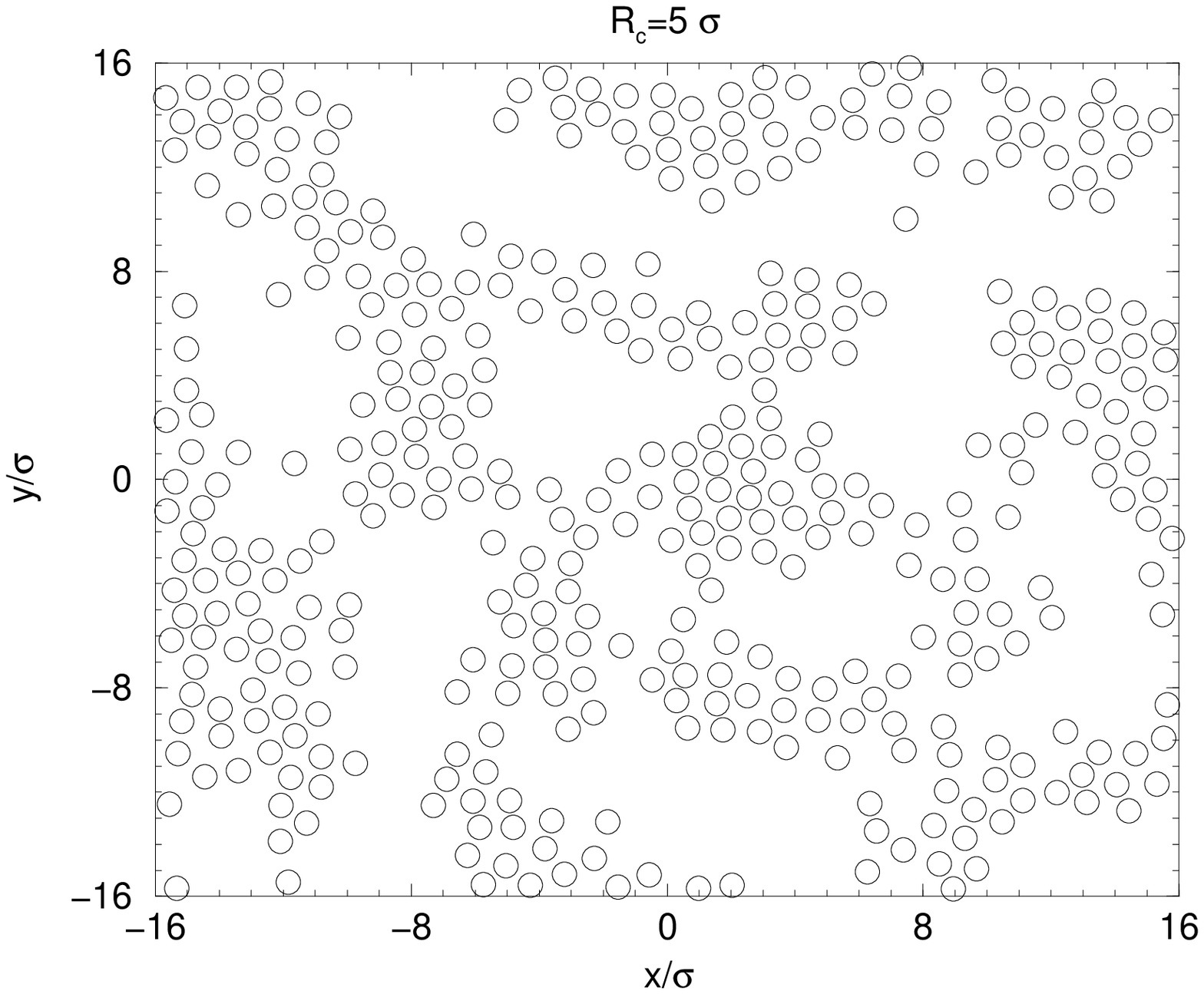}{8cm}{8cm}{0}
\mfig{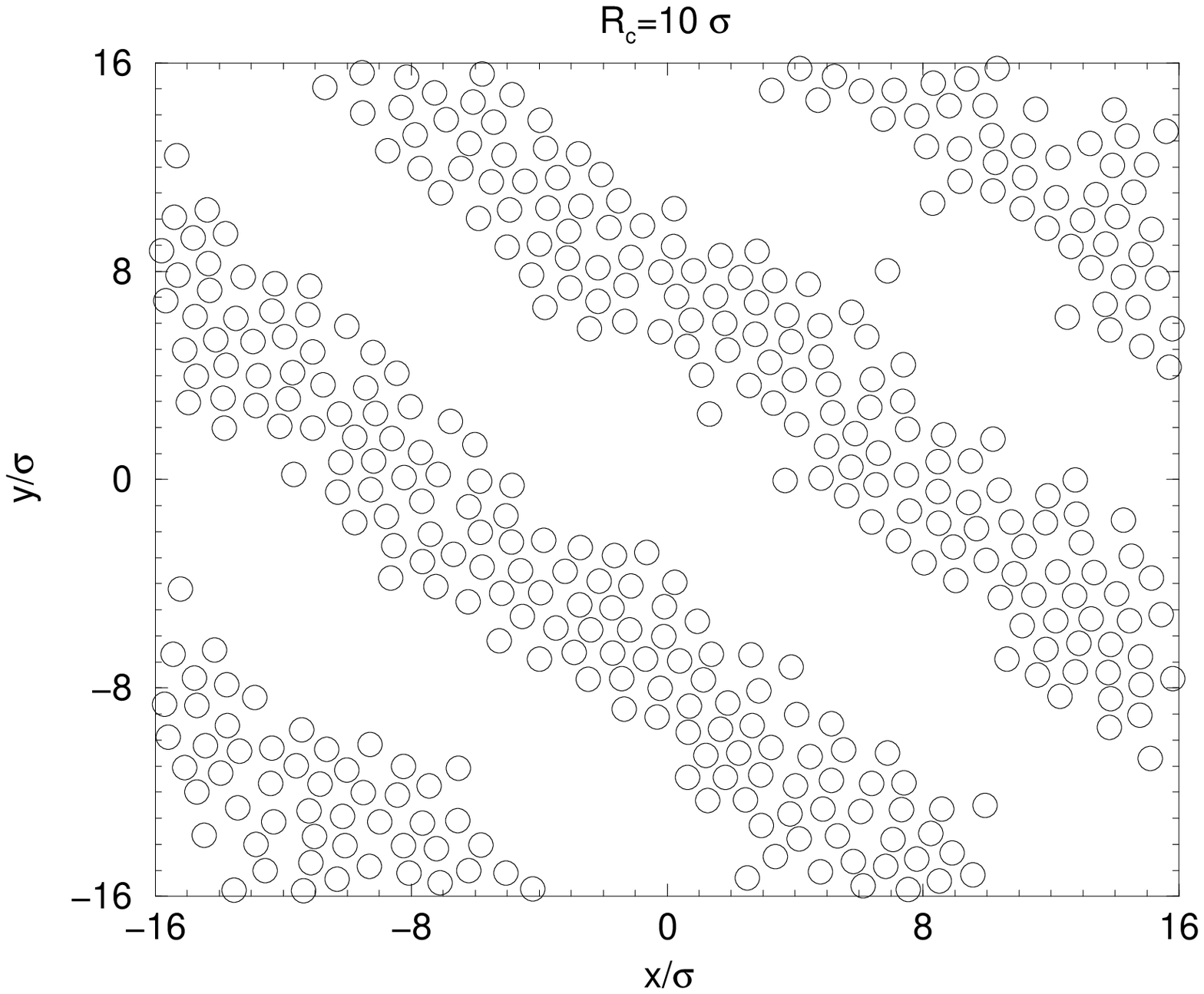}{8cm}{8cm}{0}\mfig{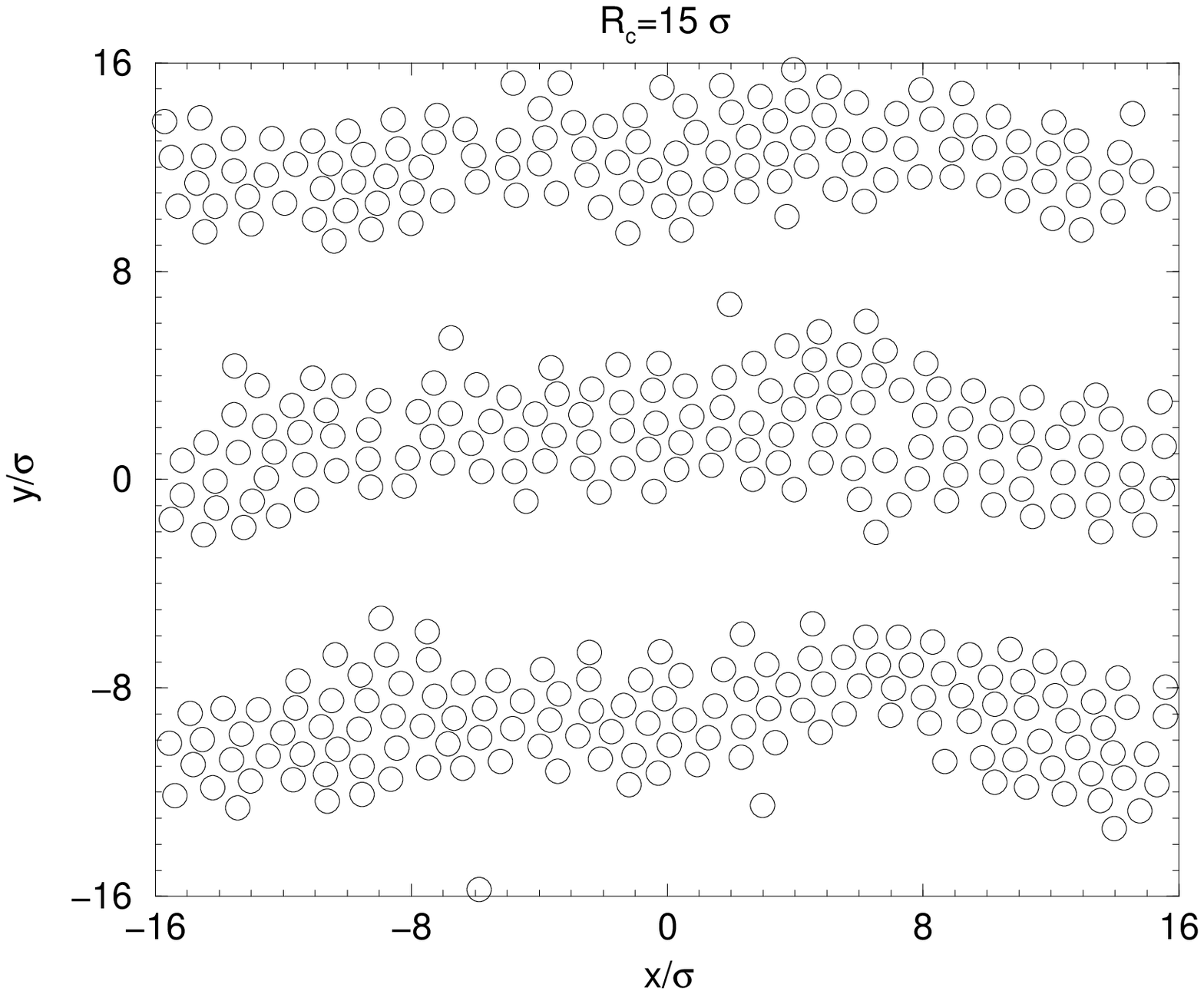}{8cm}{8cm}{0}
\end{center}
\vspace{-1cm}
\caption{\footnotesize{Snapshots relative to the state $\rho=0.4$ $T=0.5$ with different cutoff $R_c$.}}\label{ss_Rc}\end{figure}

Since we are dealing with systems subject rather to long-range interactions, we have tested different cutoff values inside the simulations. In particular, using a cutoff too short, we can completely miss the features of the state, simply because the long range repulsion becomes too weak or it is even absent. In particular at $R_c=2.5\sigma$ (so that the particles do not feel at all the repulsive hump of $U_{lr}$) the system displays a standard liquid-vapor phase transition when $T$ is low; the same happens at $R_c=3\sigma$; at $R_c=5\sigma$ we can observe the formation of domains whose shape is mostly irregular, while at $R_c=10\sigma$ and $R_c=15\sigma$ striped patterns have formed the shape of which is very regular (see figure \ref{ss_Rc}).
About the last two cases we note that an increase of the $50\%$ of the cutoff has produced an increase of $\sim 6\%$ in the pattern period. Such feature, however, does not affect the overall picture that can be traced about such phases by the present study, which is mainly aimed to characterize general behavior relative to the pattern formation.\\
Ergodicity problems were checked by starting test runs from completely different initial conditions: we have used both homogeneous configurations (with particles set on a square lattice covering the entire simulation box) and totally segregated configurations (in which all the particles belong to the same liquid-like cluster). In all the cases simulations converged to the same state.\\
The standard MC method follows a Markov chain through the configuration space. Each MC move consists of a single particle displacement and the move is accepted with probability 1 if the resulting energy change $\Delta E \leq 0$, and accepted with probability $\exp (-\Delta E/kT)$ otherwise. This method produces the correct equilibrium ensemble probability in the limit of infinite simulation time. Inside this scheme the passage of particles from a cluster to an other is not very frequent when $T$ is very small, so that the system remains ``frozen'' into a certain configuration for a long time (i.e. for many MC steps per particle).
\begin{figure}[h!]
\begin{center}
\mfig{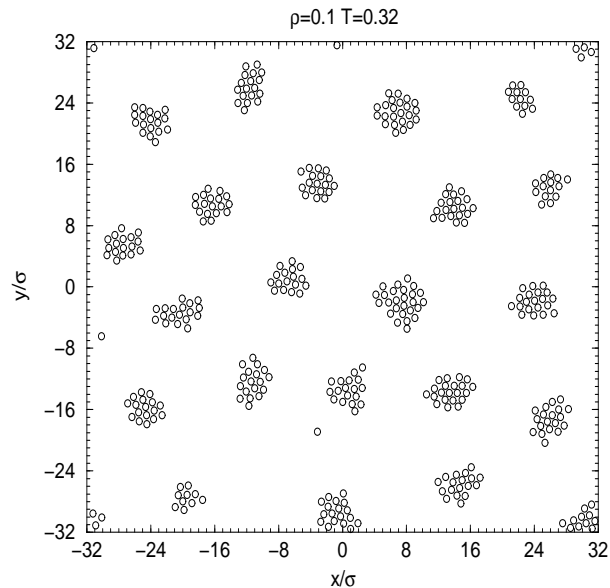}{8cm}{8cm}{0}
\end{center}
\vspace{-1cm}
\caption{\footnotesize{An example of a droplets pattern}}\label{bubble}
\end{figure}
\begin{figure}[h!]
\begin{center}
\mfig{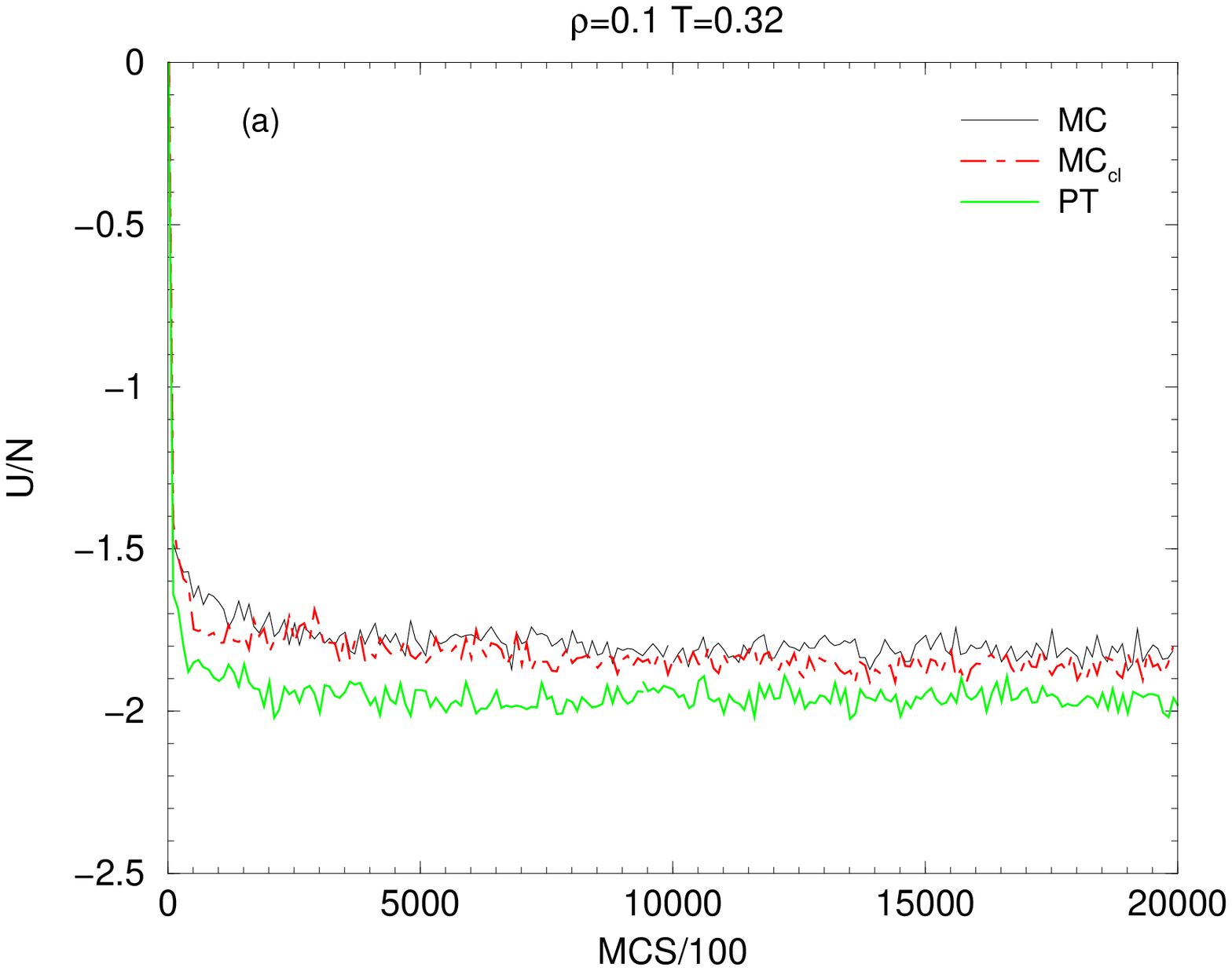}{8cm}{8cm}{0}\mfig{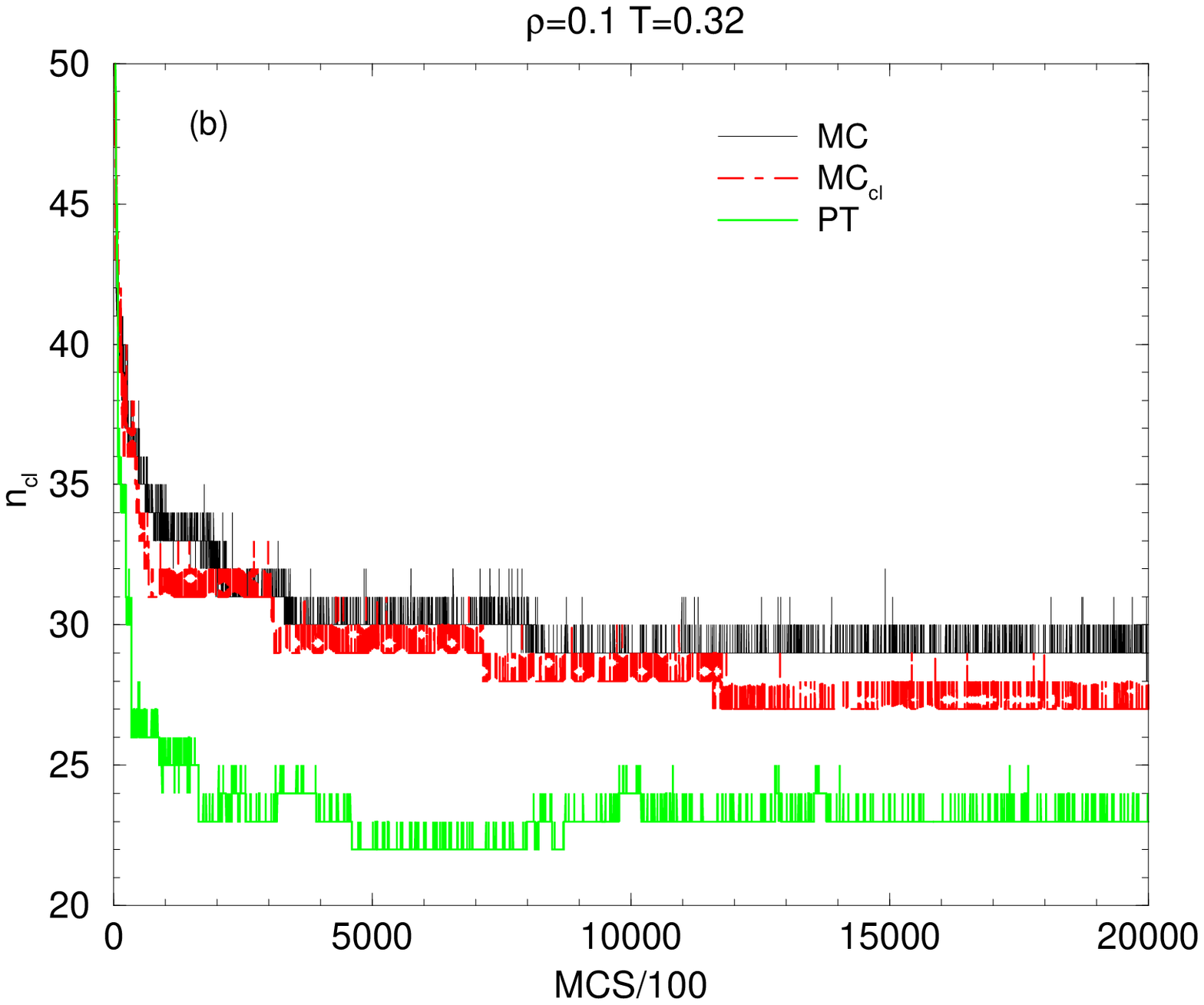}{8cm}{8cm}{0}
\end{center}
\vspace{-1cm}
\caption{\footnotesize{(a): instantaneous potential energy per particle; (b): number of clusters; the attempted cluster moves are a $20\%$ of the total MC moves; the attempted swap moves in PT are $\sim0.05\%$ of the total MC moves. MCS stands for MC steps per particle. The initial configuration was a homogeneous square lattice.}}\label{instant}
\end{figure}
We supplemented this conventional method with additional MC moves designed to accelerate the motion through the configuration space. Thus we introduced a collective move (or clusters' move) that tries to update simultaneously the position of many particles. In the following we will refer to this technique as MC$_{cl}$. At this point we want to emphasize that, here and in the following, the term cluster will be used to identify a generic collection of particles irrespective of its morphology. When we intend to refer to a particular cluster morphology we will explicitly speak of circular domains (or droplets) and stripes.\\
Into the MC$_{cl}$ scheme we choose to identify the clusters using a geometrical criterion: particles $i$ and $j$ belong to the same cluster if their separation  $d=|\protect{\bf{r}}_i-\protect{\bf{r}}_j|\leq R_o$. Typically $R_o=1.5\sigma$, corresponding to an attractive interaction between particles. In other words, the particles belonging to the same cluster are connected by a percolative path. Having identified the clusters, we must ensure that any cluster move will not prevent finding the inverse move at a later time, which means that cluster moves must not change the particle connectivity. This is achieved by displacing the center of mass of each cluster uniformly in a box of fixed side, but forbidding cluster attachment to preserve detailed balance \cite{tavares02,weis03}. Inside this scheme, in which we use the detailed balance principle and assume a symmetric transition matrix \cite{bibbia} for the probability of performing a trial move, the acceptance criterion reduces once again to $min(1,\exp(-\Delta E))$, where $\Delta E$ is the energy change between the new and the old configurations. The collective moves represent the $20\%$ of the total.\\
In this way large changes of the system can be obtained in a single MC step. Such moves are favourable mainly at low temperature where the mean displacement  of a cluster  can be $1-2$ order of magnitude bigger than the single particle one. At high temperature this technique is not so effective because the greater probability of clusters' overlapping.
 This scheme enhances the sampling of the configuration space with respect to the simple MC technique.\\
In addition, to be sure to sample sufficiently the configuration space and that we were not trapped into a local minimum of the free energy landscape, we adopted the Parallel Tempering algorithm (PT) (\cite{bibbia,opps01} \cite{faller02} and references therein). The PT technique was developed for dealing with the slow dynamics of disordered spin systems. The PT algorithm simultaneously simulates a set of $M$ identical non interacting replicas of a system, each of them at a different temperature. Periodically a swap between configurations belonging to different replicas is attempted. The exchange is accepted in a Metropolis fashion applied to an extended ensemble which is a combination of subsystems that are $NVT_i$ ensembles \cite{bibbia}. The acceptance criterion is:
\begin{equation}
W_{m,n}=\left\{\begin{array}{ll}
                 1 & \Delta_{m,n}\leq0\\
		 \exp(-\Delta_{m,n}) & \Delta_{m,n}>0
		 \end{array}
        \right.
\end{equation} 
where $\Delta_{m,n}=(\beta_n-\beta_m)(E_m-E_n)$, with $E_m$ $E_n$ energies of the replicas of index $m$ $n$ and $\beta$ is the inverse of temperature. The success of PT relies on the fact that the temperature range covers values high enough to pass every free energy barrier. The outcome of such calculations is, in principle, equilibrium configurations in the canonical ensemble at the $M$ different temperatures. In our simulation we took $M=31$ and we covered the range $0.32\leq T\leq 0.92$. Each replica evolved according to a MC$_{cl}$ scheme with $0.05\%$ attempted swap moves on the total MC moves. Exchange events were examined only between  subsystems that are nearest neighbours in temperature: $i$ and $i+1$ \cite{yamamoto00}. Comparing PT with the techniques previously described we find that it is very efficient to improve convergence and  sample equilibrium configurations, specially for the low temperature states. See, for example, figure \ref{instant} where we  plotted the number of clusters and the potential energy per particle for the thermodynamic state $\rho=0.1$ and $T=0.32$ (a snapshot of which is shown in figure \ref{bubble}).
\begin{figure}[h!]
\begin{center}
\mfig{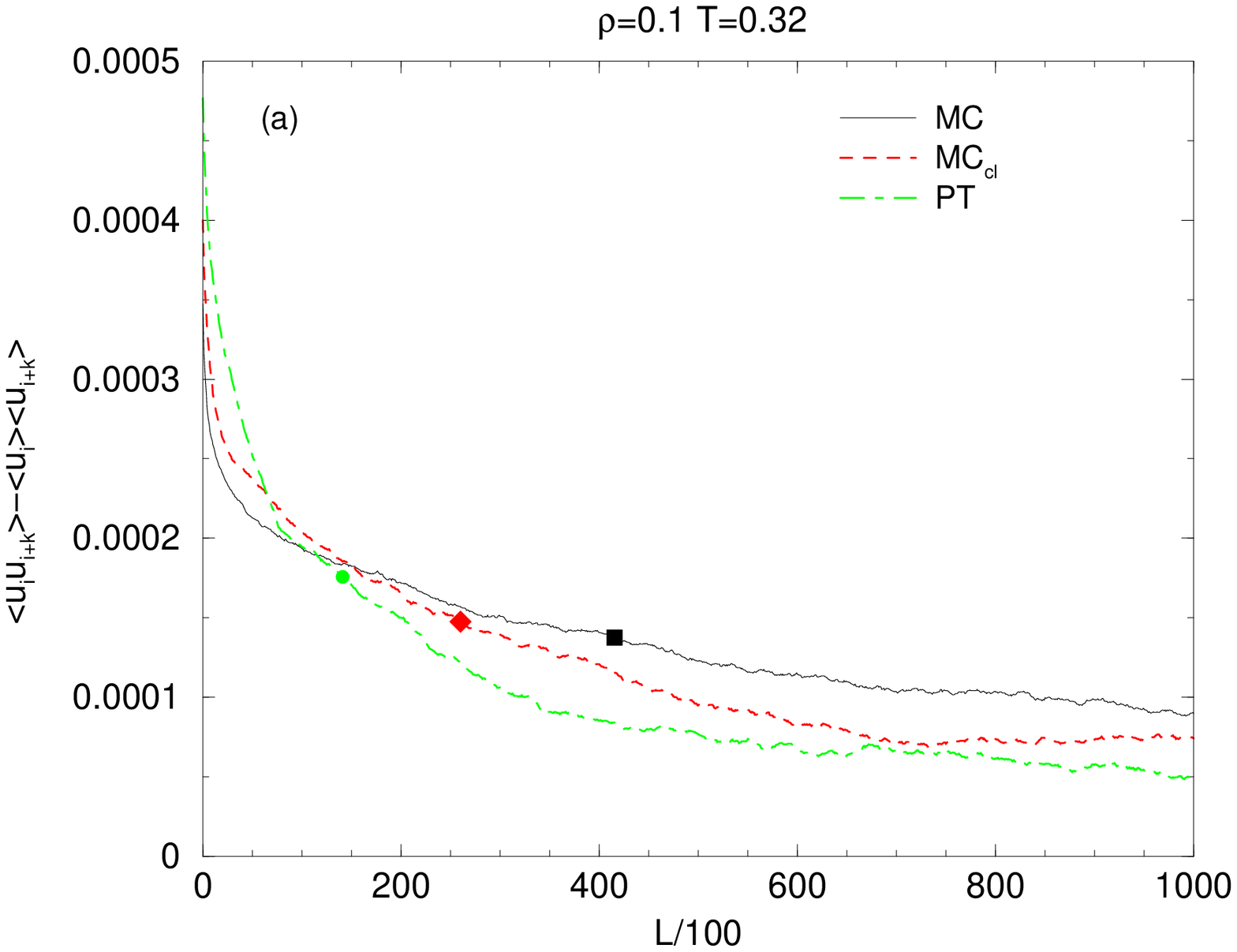}{8cm}{8cm}{0}\mfig{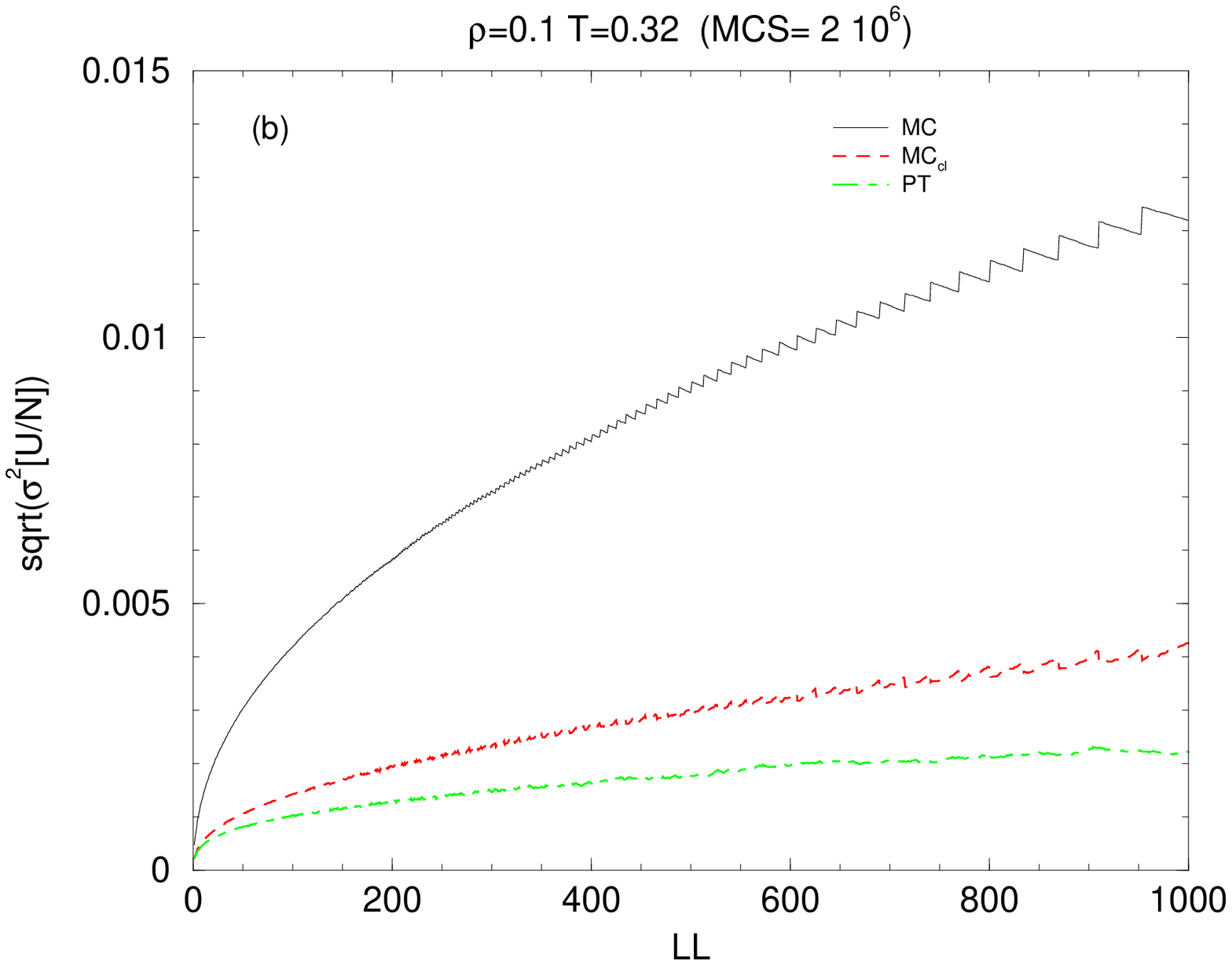}{8cm}{8cm}{0}
\end{center}
\vspace{-1cm}
\caption{\footnotesize{(a): covariance versus distance in MCS (MC steps per particle) between configurations generated with different MC schemes; $u=(U/N)$. Filled symbols: $1/e$ decay (read text) with respect to the initial value. (b): estimator of variance versus the length of blocks used into data blocking scheme for computation of statistical errors.; LL is the number of configurations into each block. }}
\label{corr}
\end{figure}
In figure \ref{corr}(a) we plotted the covariance relative to $U/N$ obtained with different simulation schemes versus the distance ( expressed in MC steps) between two different configurations. On each line the filled symbols mark the point where the covariance drops by a factor $1/e$ with respect to the initial value: the configurations generated with the PT algorithm become uncorrelated faster than they do with the other methods. In figure \ref{corr}(b) we plotted the estimators for the variance, using the data blocking technique, for the computation of statistical averages and errors: if the simulation is sufficiently long the estimator tends toward a constant \cite{bibbia}. We see that for a simulation run of fixed length, the PT curve is the flattest one; i.e., using PT technique we need shorter runs. A typical length of our PT runs is $\sim 5\cdot 10^6$ MC steps per particle.

\section{Simulation results\label{Microphases: simulation results}}\begin{figure}[h!]
\begin{center}
\mfig{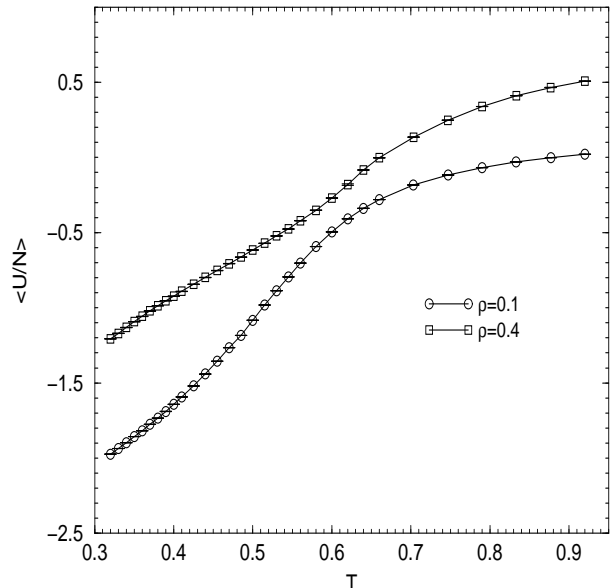}{8cm}{8cm}{0}
\end{center}
\vspace{-1cm}
\caption{\footnotesize{Simulation data for the average excess internal energy per particle at different densities. The statistical uncertainty is smaller than the symbols size. Continuous lines are only guide to eyes.}}\label{Uconf}
\end{figure}
\begin{figure}[h!]
\begin{center}
\mfig{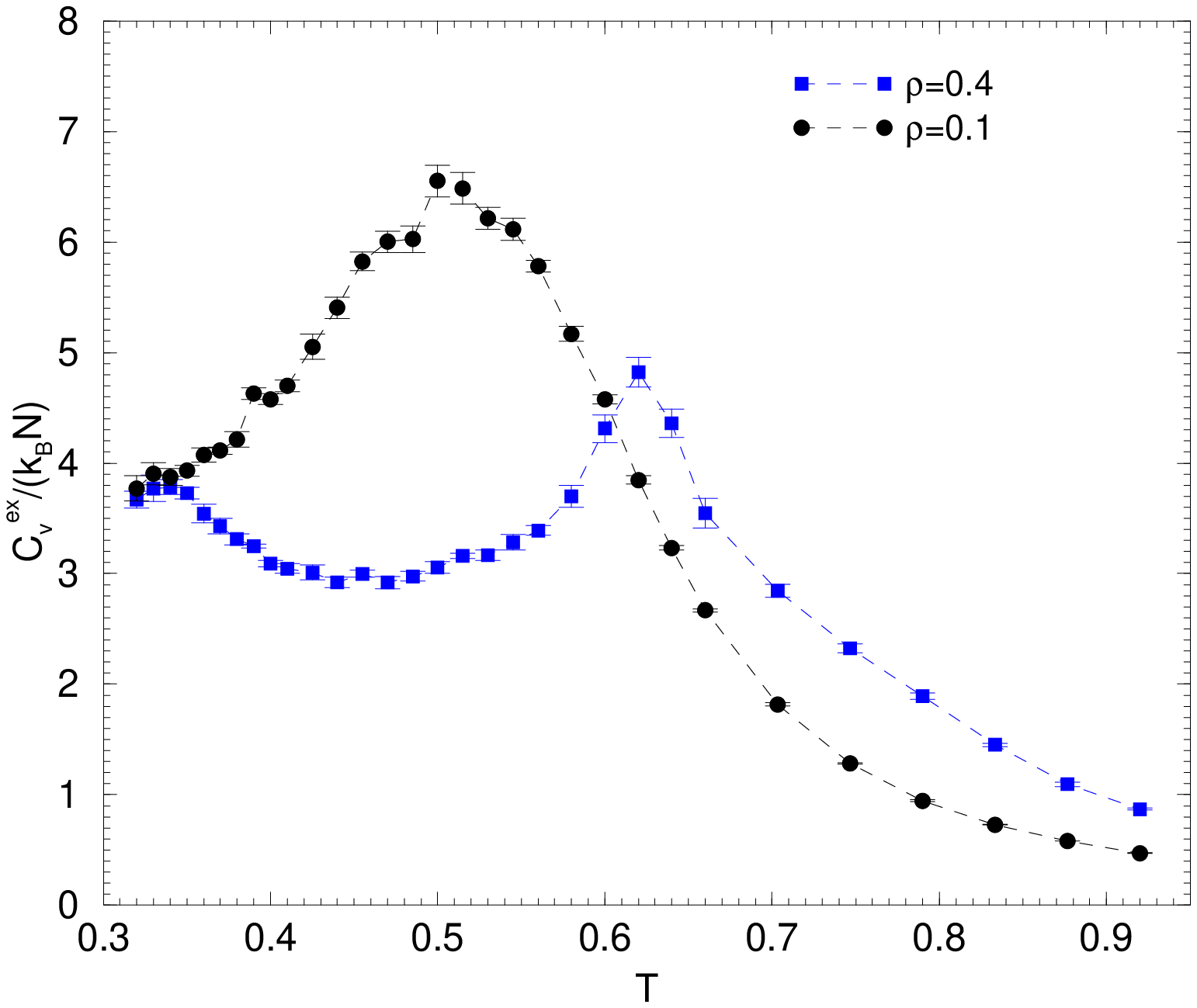}{8cm}{5cm}{0}

\mfig{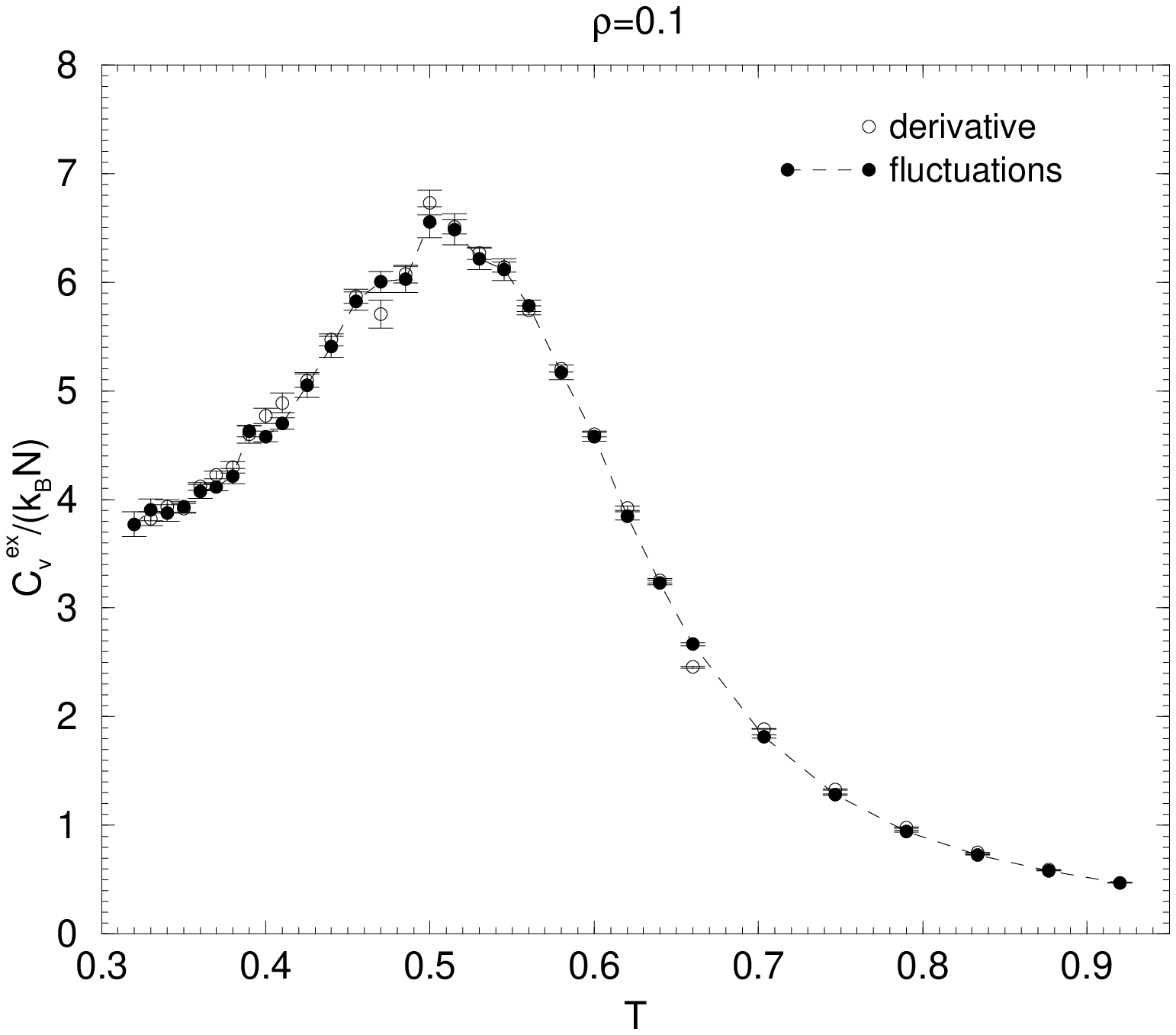}{8cm}{5cm}{0}

\mfig{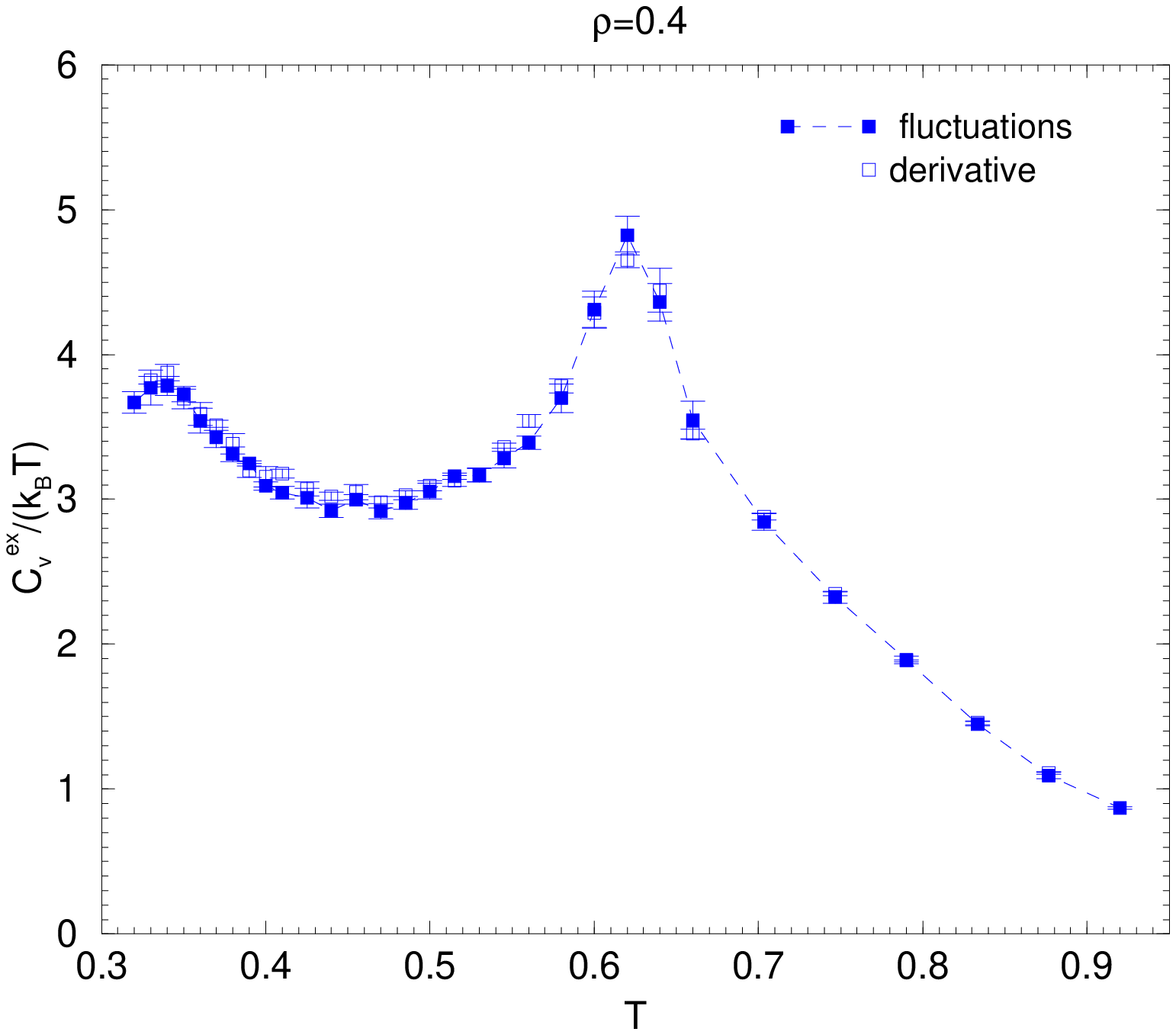}{8cm}{5cm}{0}
\end{center}
\vspace{-1cm}
\caption{\footnotesize{Upper panel: comparison of the dimensionless excess heat capacity per particle for densities $\rho=0.1$ and $\rho=0.4$ computed according to the equation \ref{fluctuations}; middle and bottom panels: comparison between the two methods of computing $C_v^{ex}/(k_BT)$ according to the equations \ref{fluctuations} and \ref{derivative}. Dashed lines are only guide to eyes.}}\label{Cv}
\end{figure}
\begin{figure}[h!]
\begin{center}
\mfig{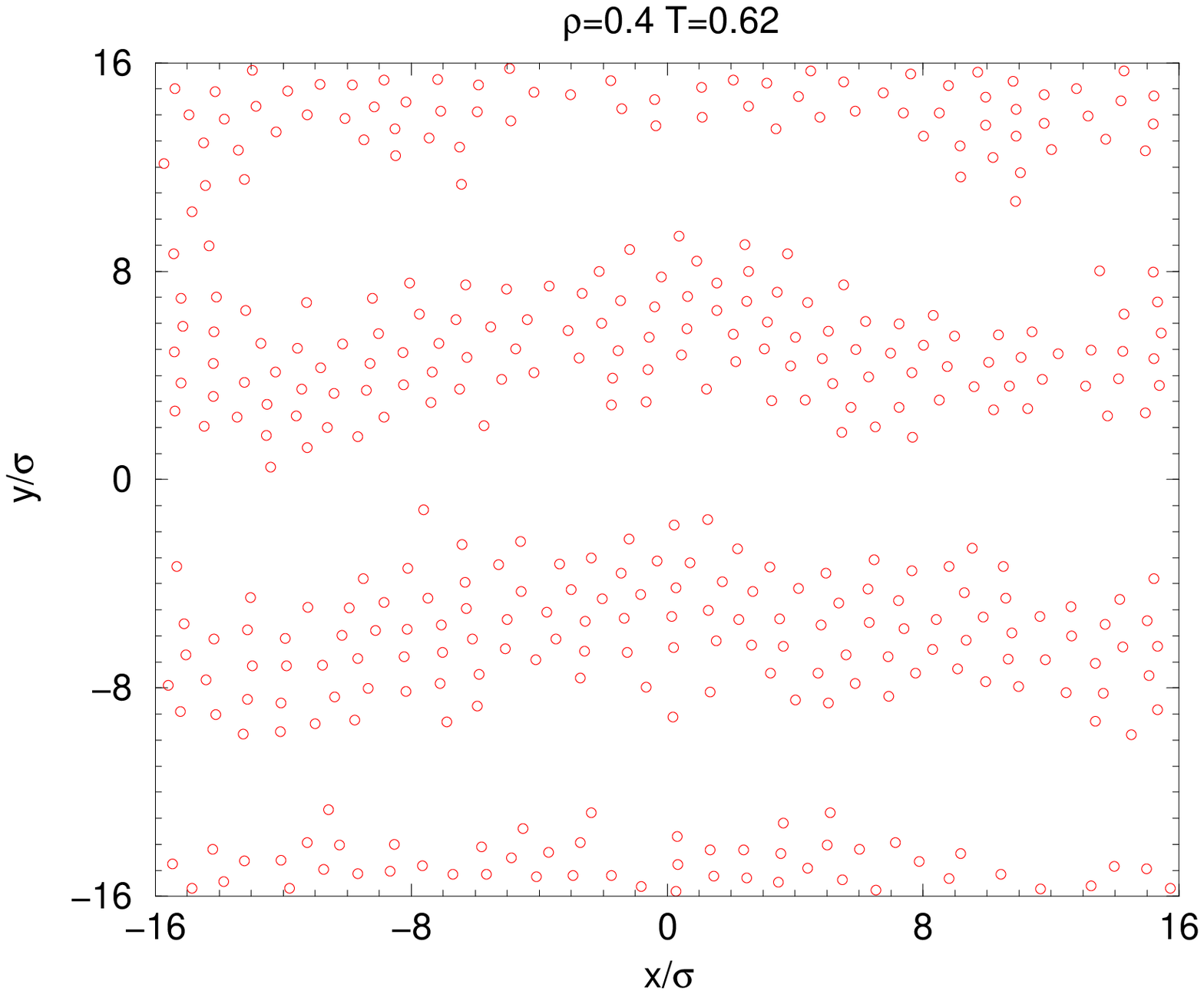}{8cm}{8cm}{0}\mfig{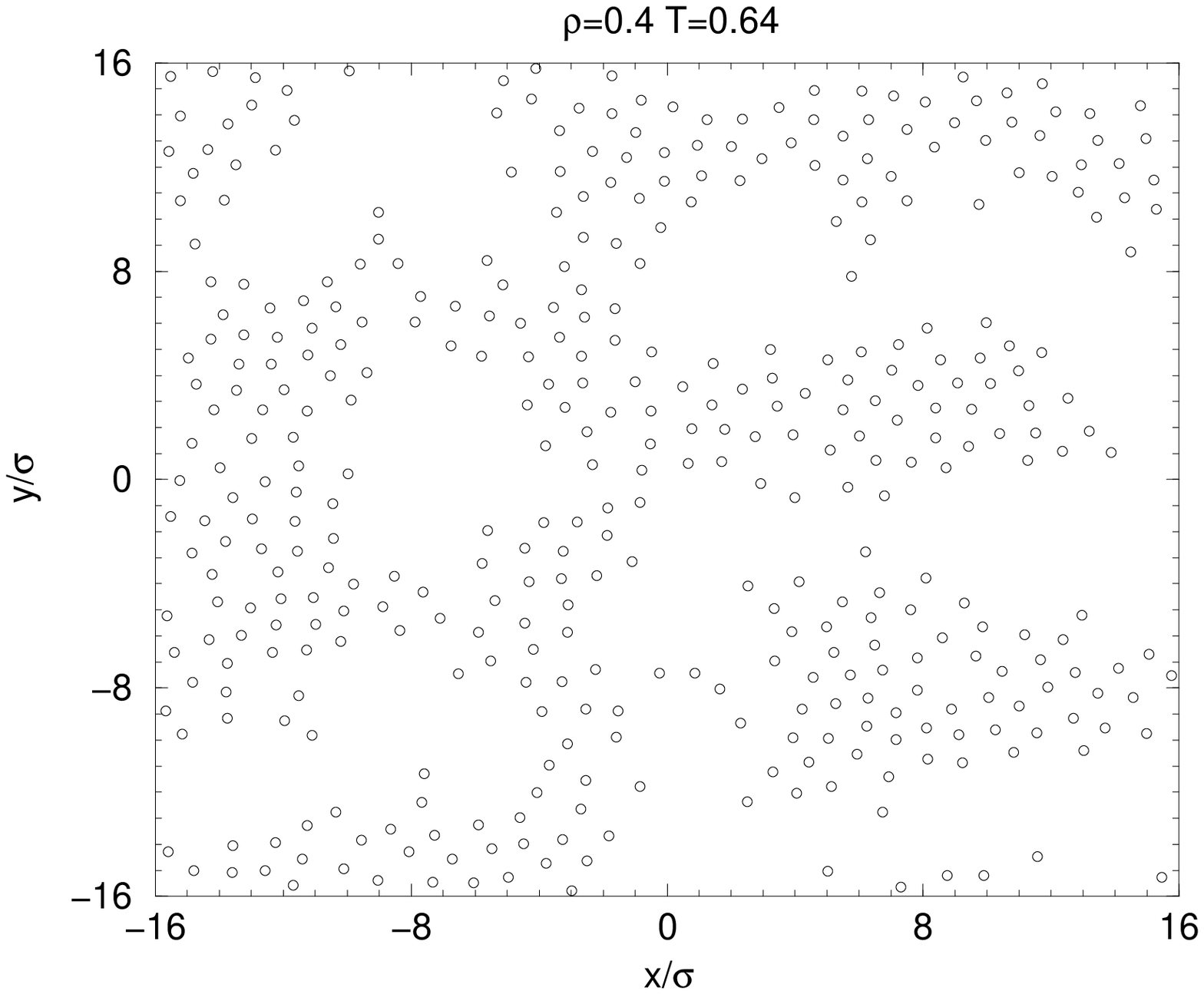}{8cm}{8cm}{0}
\end{center}
\vspace{-1cm}
\caption{\footnotesize{Snapshots relative to the microphases-homogeneous fluid transition.}}\label{ss}
\end{figure}
\begin{figure}[h!]
\begin{center}
\mfig{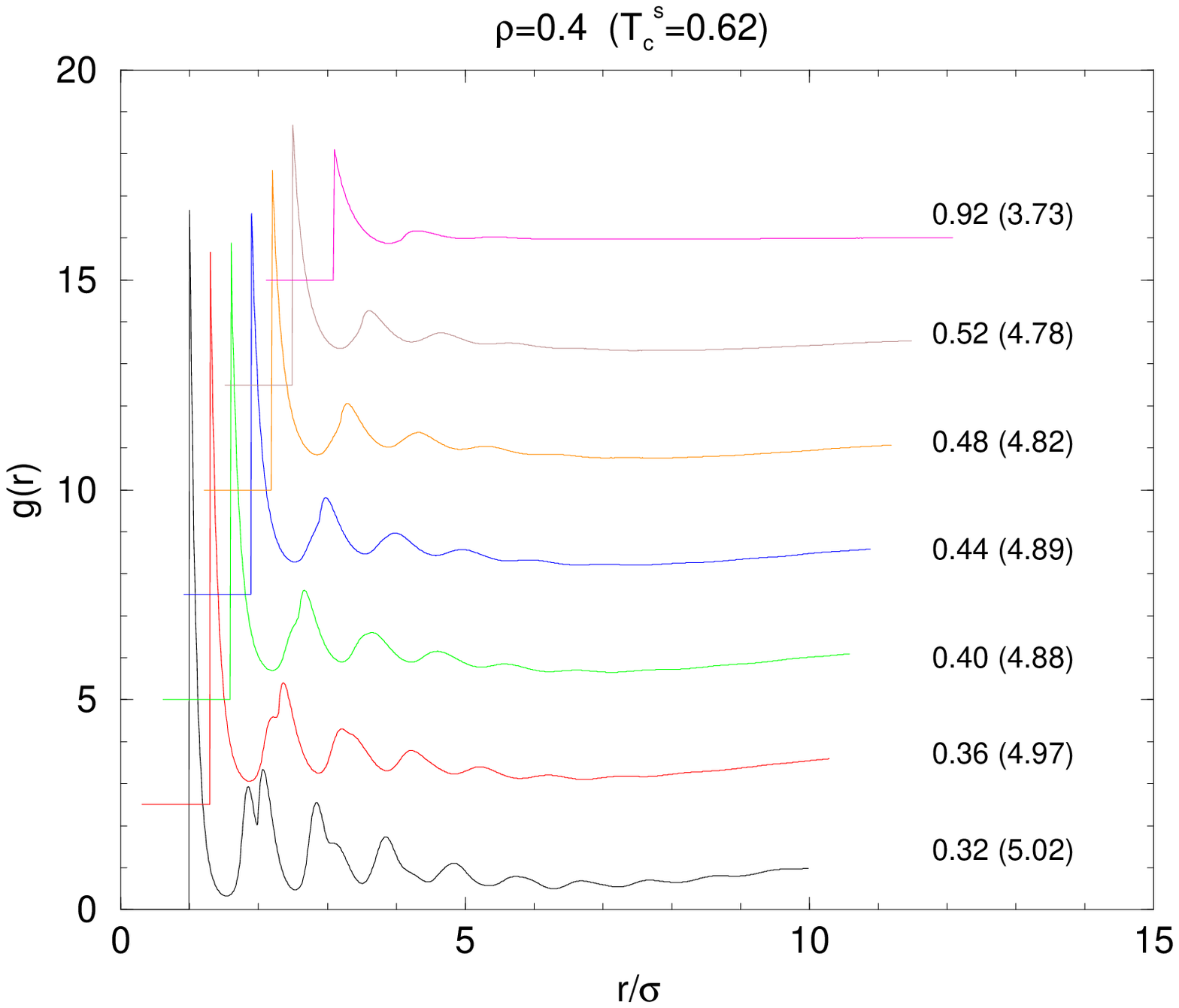}{8cm}{8cm}{0}\mfig{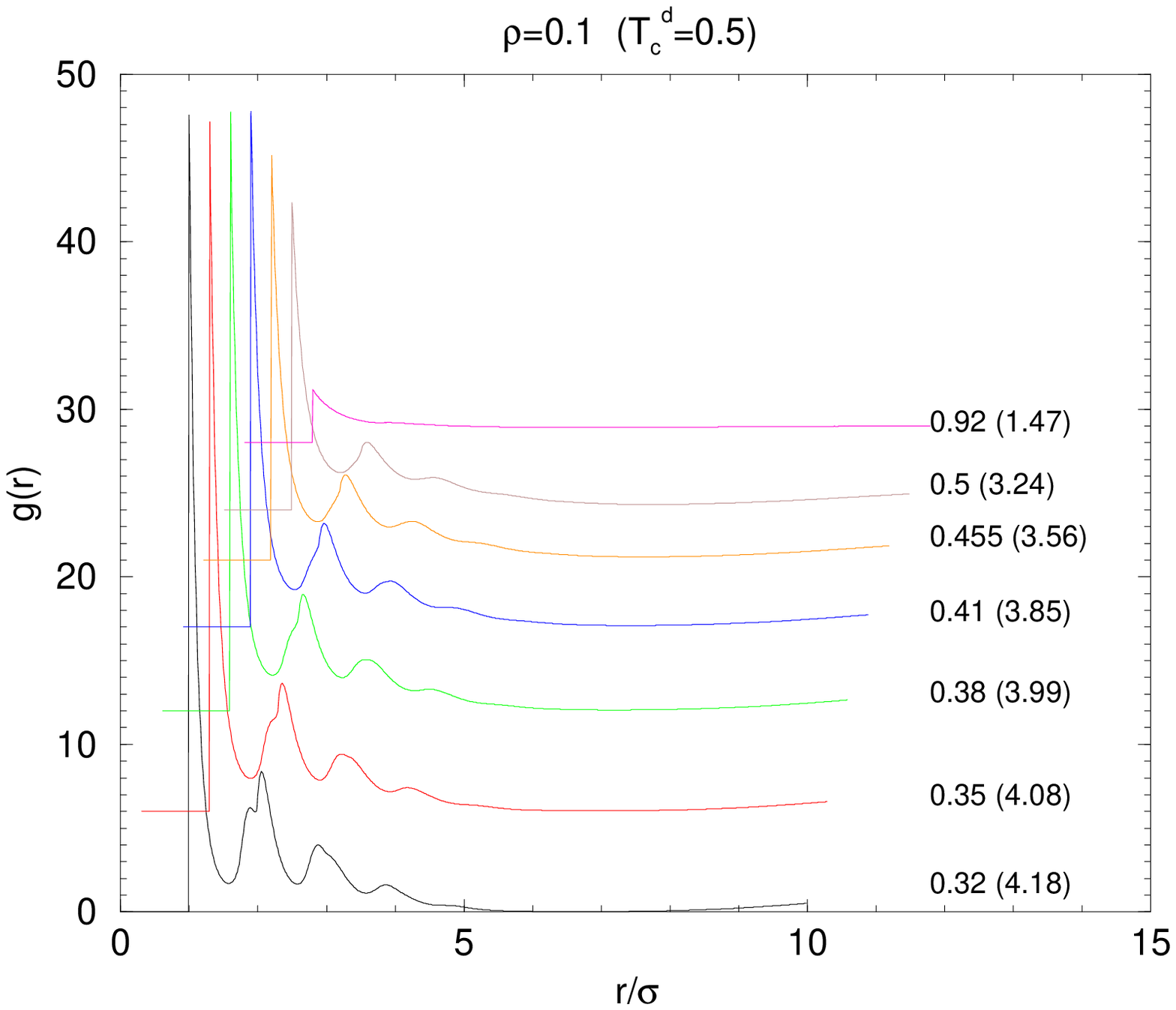}{8cm}{8cm}{0}
\end{center}
\vspace{-1cm}
\caption{\footnotesize{Left: $g(r)$ for $\rho=0.4$; right: $g(r)$ for $\rho=0.1$; next to the curves the temperature and (into parenthesis) the coordination number are indicated; the curves are shifted for clarity. The critical temperature is  also reported.}}\label{G}
\end{figure}
\begin{figure}[h!]
\begin{center}
\mfig{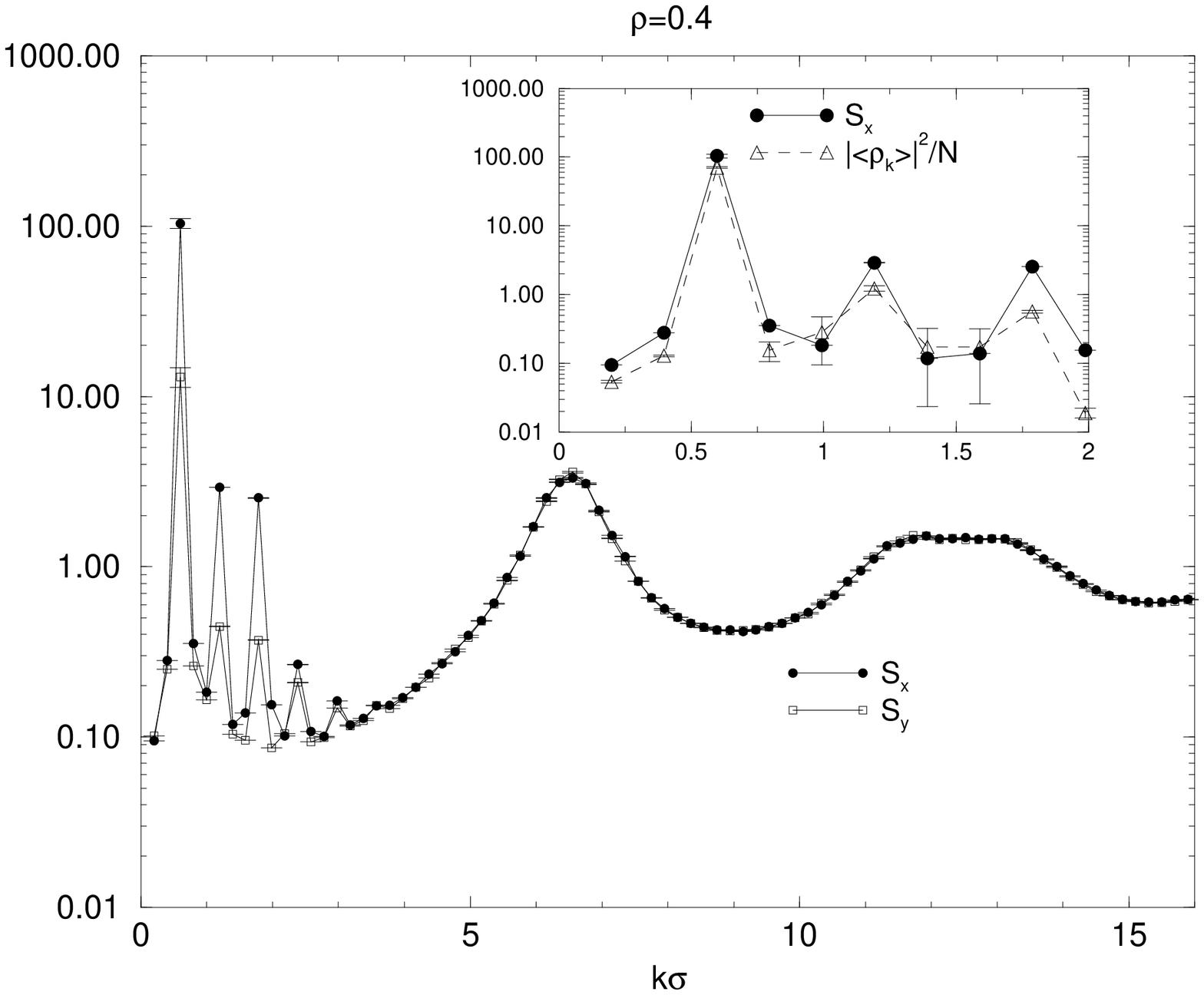}{8cm}{8cm}{0}\mfig{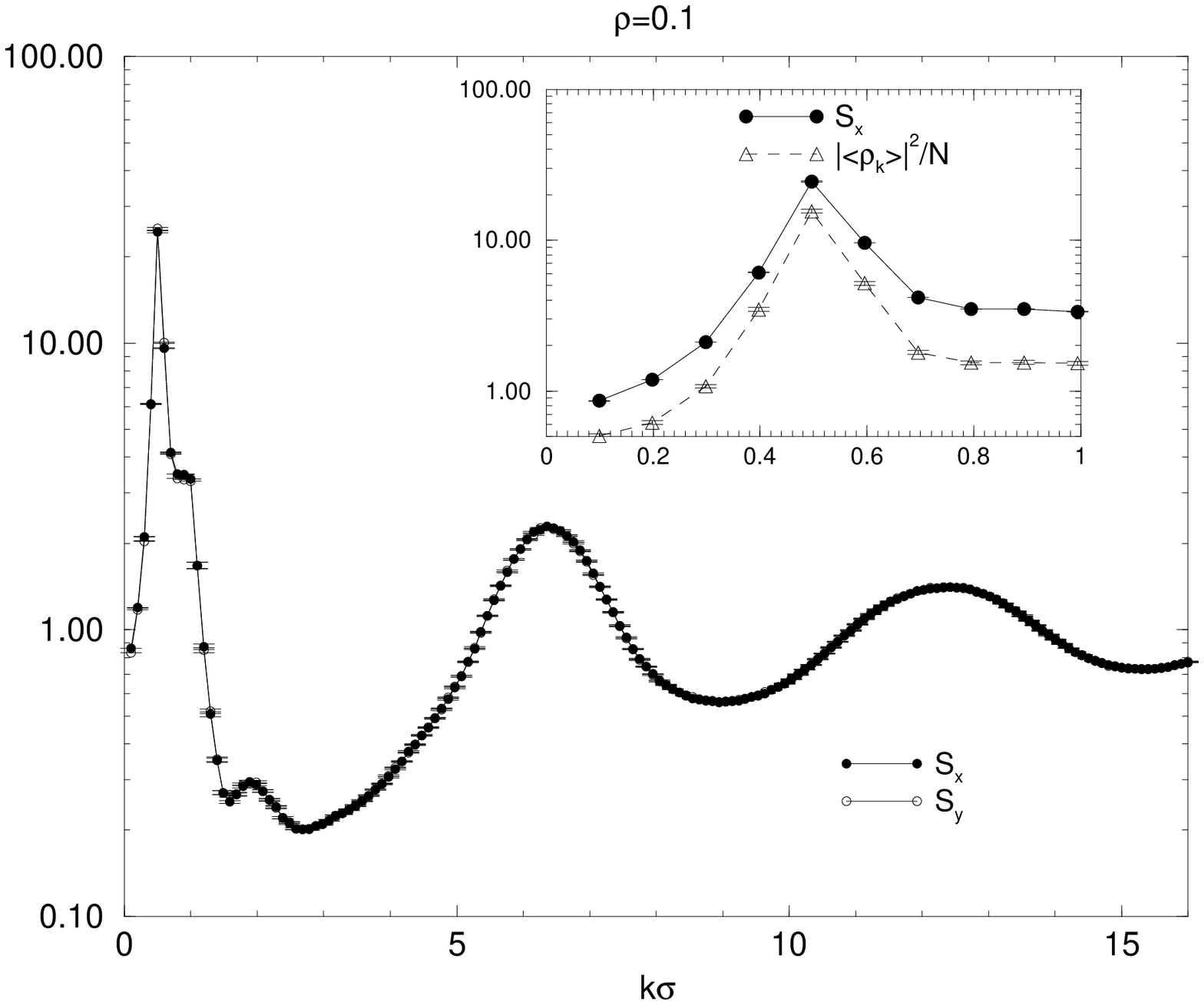}{8cm}{8cm}{0}
\end{center}
\vspace{-1cm}
\caption{\footnotesize{Static structure factor at $T=0.36$ computed along $x$ and $y$ directions for different densities. The insets show the structure factor compared with an estimator of the Fourier transform of the density profile along the same direction.}}\label{S}
\end{figure}

We report our results for the interaction potential  of equation \ref{kac}; the parameters of the potential are $\epsilon_a=\epsilon_r=1$, while the ranges are $R_a=1\sigma$ and $R_r=2\sigma$. Initial investigations were done using the standard MC algorithm, from which we located the region of phase diagram corresponding to different pattern formation. We explored the density range $\rho= (0.1\div0.8)$ and the temperature range $T=(0.3\div1.0)$. At $T < 0.7$ we could observe the formation of microphases. In particular for $\rho\lesssim0.35$ circular domains of particles (i.e. droplets) were observed while for $0.35\lesssim\rho<0.6$ liquid-like striped patterns formed. This is qualitatively in agreement with many experimental results in which, as the density increases, the cluster morphology changes to form pattern with lower curvature \cite{sear99,gelbart99}. At higher densities gas bubbles inside a liquid medium could be observed as a counterpart of the liquid droplets. The aim of the present computations was to characterize only the major features of the phase diagram, so that we cannot exclude the presence of some additional morphology mainly in the transition region between droplets and stripes. The characteristic dimension of the pattern is simply connected to the range of the potential: the droplet diameter as well as the stripe width is $\sim (4\div 5) \sigma$ that roughly corresponds to the location of the repulsive hump of $U_{lr}$. Extensive computations were done with the PT and the cluster moves technique which essentially confirmed the preliminary runs but allowed more precise computation of the thermodynamic and structural quantities of the system. In this paper we report the results relative to two particular densities $\rho=0.1$ and $\rho=0.4$ in which the system exhibits droplet and striped pattern respectively. The simulation results shown in this section are obtained adopting the PT if not otherwise specified.\\
In figure \ref{Uconf} we plotted the average excess internal energy per particle $<U/N>$ for the droplet and stripe cases: at higher density the excess internal energy  is less than it is in the low density case because particles, which belong to stripes at $\rho=0.4$, feel a stronger reciprocal repulsion than they do when they are set into droplets at $\rho=0.1$. In the following section we will discuss, within a simple model, the change of cluster morphology as the density increases.\\ 
In order to identify the phase transition leading to the loss of order of the microphase region, we computed the dimensionless excess heat capacity  per particle as shown in figure \ref{Cv}. We have computed in two ways $C_v^{ex}/Nk_B$ at constant volume:
\begin{equation}
\frac{C_v^{ex}}{Nk_B}=\frac{1}{Nk_B}\frac{d<U>}{dT}\label{derivative}
\end{equation}
\begin{equation}
\frac{C_v^{ex}}{Nk_B}=\frac{1}{N}\frac{<U^2>-<U>^2}{(k_BT)^2}\label{fluctuations}
\end{equation}
Since we study the system at discrete temperatures, the derivative is approximated with finite differences so that:
\begin{equation}
\frac{C_v^{ex}(T_n)}{Nk_B}=\frac{1}{Nk_B}\frac{<U(T_{n+1})>-<U(T_{n-1})>}{T_{n+1}-T_{n-1}}
\end{equation}
with $T_{n+1}>T_{n-1}$.
In equilibrium these two ways of calculating $C_v^{ex}$ should agree.
In the top panel of figure \ref{Cv} we have shown the results relative to the second method, while a comparison between the two methods for each density is shown in the middle and bottom panels. The agreement between the two methods is quite good, suggesting that the systems have well equilibrated at all temperatures. The peaks of the specific heat identify the phase transition from the microseparated region to the homogeneous one. The corresponding transition temperatures are $T_c^d=0.5$ for the droplets and $T_c^s=0.62$ for the stripes. The latter case shows a particularly pronounced peak and also the snapshots of the system confirm an abrupt passage from an ordered configuration to a disordered one (see figure \ref{ss}).
It is interesting to note that the specific heat for $\rho=0.4$ has a maximum at $T=0.34$ and this seems to be connected to the freezing transition inside the stripes, as we will discuss below treating structural quantities such as the radial distribution function  $g(r)$ and the static structure factor $S(k)$.\\
In figure \ref{G} we have plotted $g(r)$ (computed averaging over all the directions and normalized to the mean density of the system) for different temperatures; in the same figure we have indicated the coordination number $n_c$:
\begin{equation}
n_c=2\pi\rho\int_0^{R_{min}}r g(r) \rmd r,
\end{equation}
in which $R_{min}$ is the position of the first minimum.
The curves are shifted for clarity; we can identify two regimes: a short-range modulation due to the local structure of the fluid and a longer-range modulation due to the microphases formation. The latter manifests itself with a shallow minimum around $\rho \sim 6 \sigma$ at low temperature, which is doomed to disappear as the system becomes homogeneous at larger $T$.
As temperature decreases, the second peak of $g(r)$ develops a shoulder which eventually grows till to a really peak splitting. The appearance of this shoulder has been referred to as a structural freezing precursor both in two and three dimensions \cite{truskett98}. The shoulder appears at $T\sim (0.38\div0.40)$ for droplets and at $T\sim (0.44\div0.46)$ for stripes, while the splitting is clearly visible at $T=0.32$ for droplets and $T=0.38$ for stripes. The peak splitting is consistent with a triangular lattice of period $a\sim 1.05\, \sigma$.\\
The static structure factor has been calculated by explicit evaluation of the expression:
\begin{equation}
S(\protect{\bf{k}})=N^{-1}<(\sum_i^N\cos(\protect{\bf{k\cdot r}}_i))^2+(\sum_i^N\sin(\protect{\bf{k\cdot r}}_i))^2>,
\end{equation}
with $\protect{\bf{k}}$ along different directions so, in the following, we will refer to $S_x(k)$ and $S_y(k)$ as the static structure factor with $k$ along the $x$ and $y$ directions respectively. Due to the finite size of the simulation box, we can compute such quantity only for wave vectors multiple of the smallest vector $k_o=2\pi/L_b$ ($L_b$ box simulation side). In figure \ref{S} we have reported the results at one temperature for the two studied densities.
We can recognize two regimes. For $k\sigma<3$ the peaks are strictly connected to the modulation of the pattern which the clusters arrange themselves on. We refer to the peaks at low $k$ as to the modulation peaks. The height of the modulation peak strongly depends on the direction of the stripes and the direction of $\protect{\bf{k}}$ along which we are computing $S(k)$, i.e., if the stripes are close to one of the principle axis of the simulation box or not. For instance, at $\rho=0.4$ plotted in figure \ref{S}, the $x$ direction lays perpendicular to the stripes, while the $y$ direction is quite parallel to them. Into the inset of the same figure we have also shown a comparison between $S(k)$ and the square of the Fourier transform of the density modulation $|<\rho_k>^2|/N$, to underline that the  peaks to a large extent are due just to the density modulation. Of course on a very long run the pattern will fluctuate giving a uniform density so $<\rho_k>$ will vanish whereas $S(k)$ maintains the modulation peaks. The location of the main modulation peak corresponds to a modulation period $\lambda=2\pi/k\sim11\sigma$. For $k\sigma>3$, instead, the behavior of $S(k)$ is determined mainly by the structure of the fluid inside the clusters. As a matter of fact, at low temperature we can observe a flattening of the peaks at $k\sigma>10$ till to a true peak splitting (at the lowest temperatures), which is just the counterpart into the Fourier space of the peaks splitting observed into the radial distribution function.

\begin{figure}[h!]
\begin{center}
\mfig{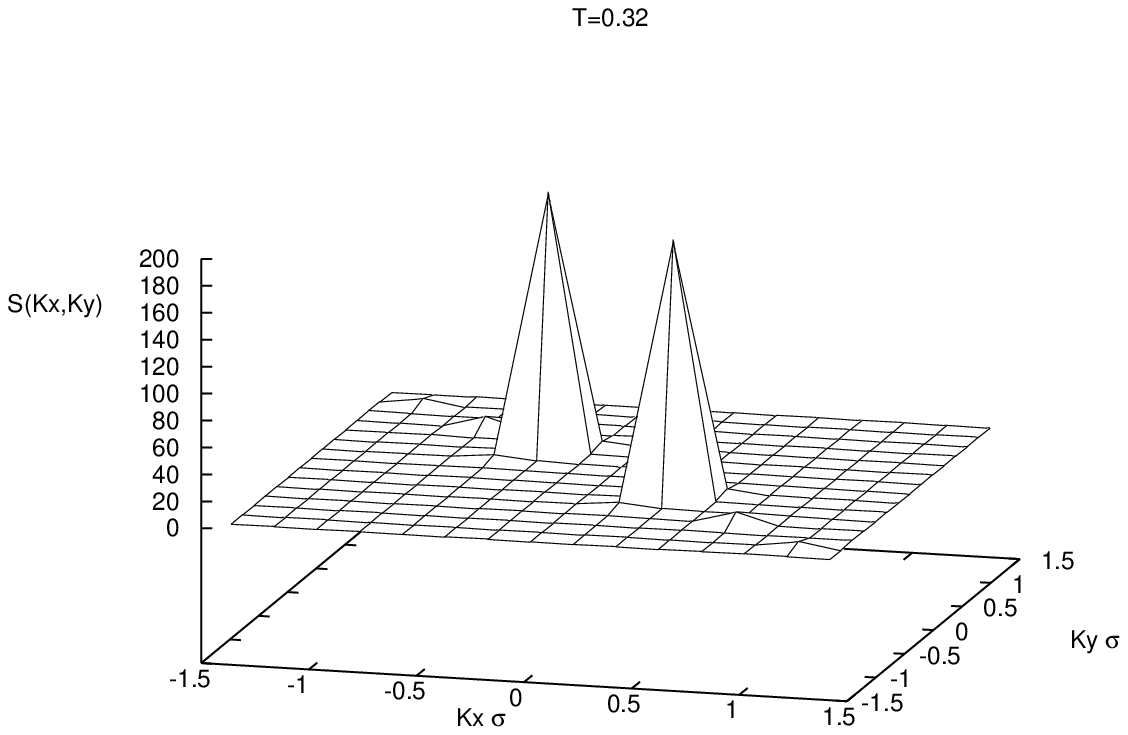}{8cm}{7cm}{0}\mfig{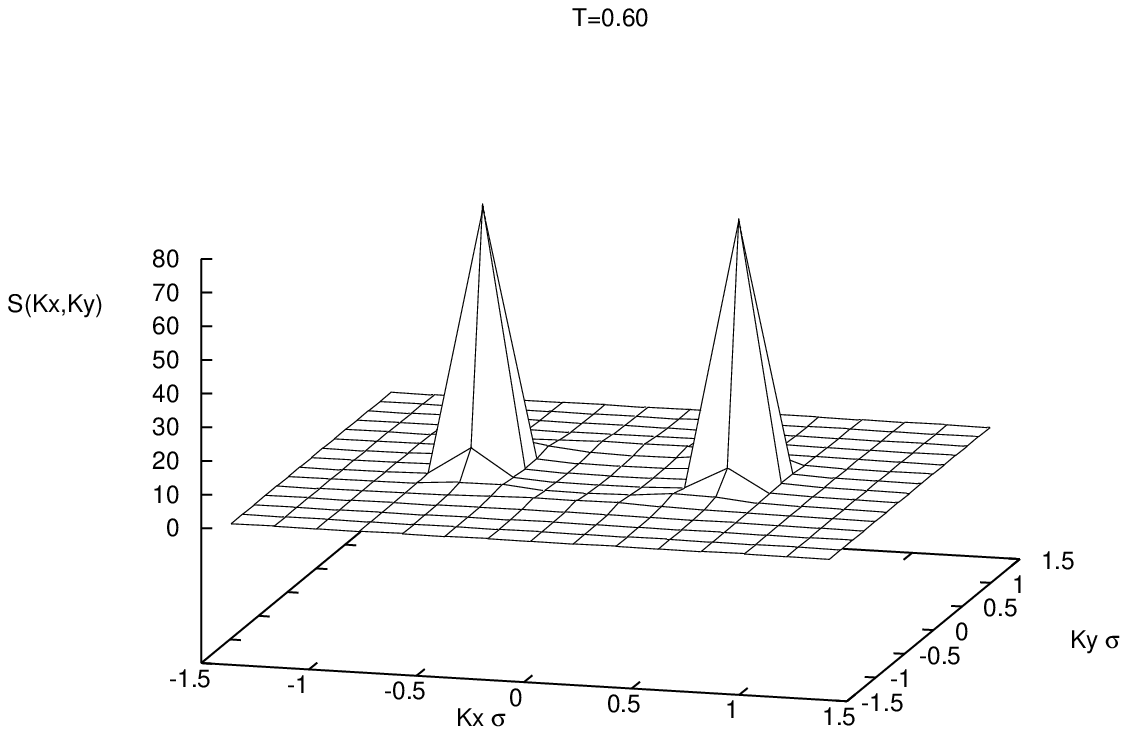}{8cm}{7cm}{0}
\mfig{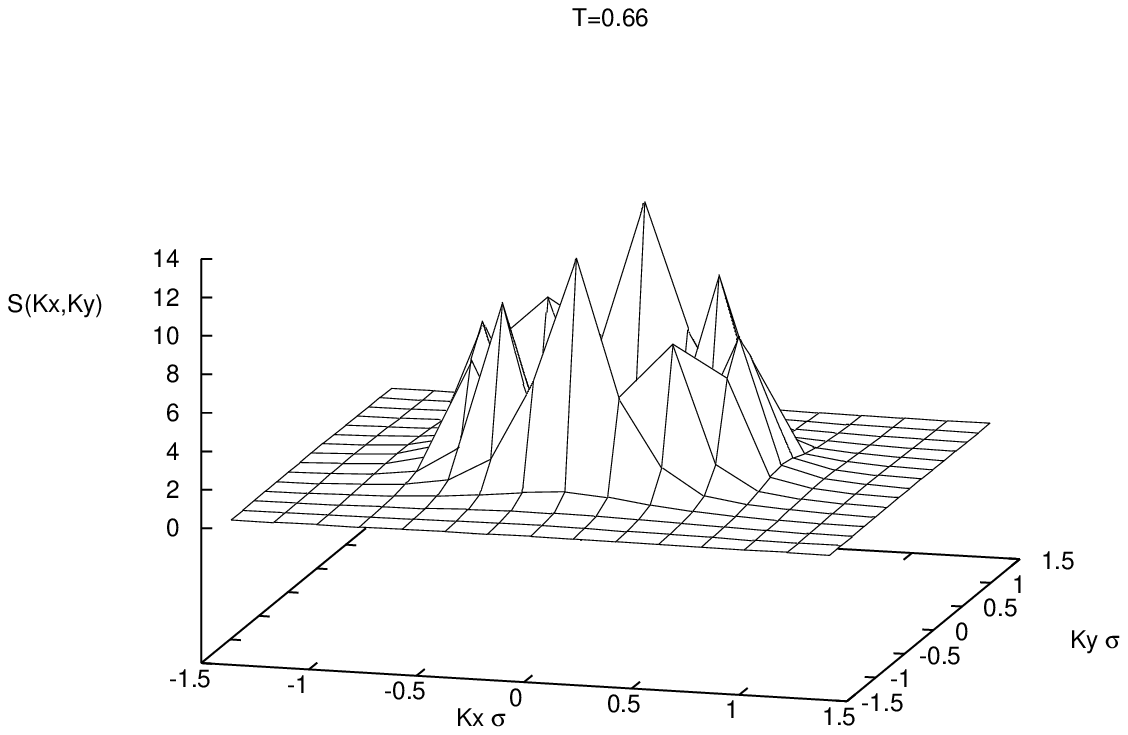}{8cm}{7cm}{0}\mfig{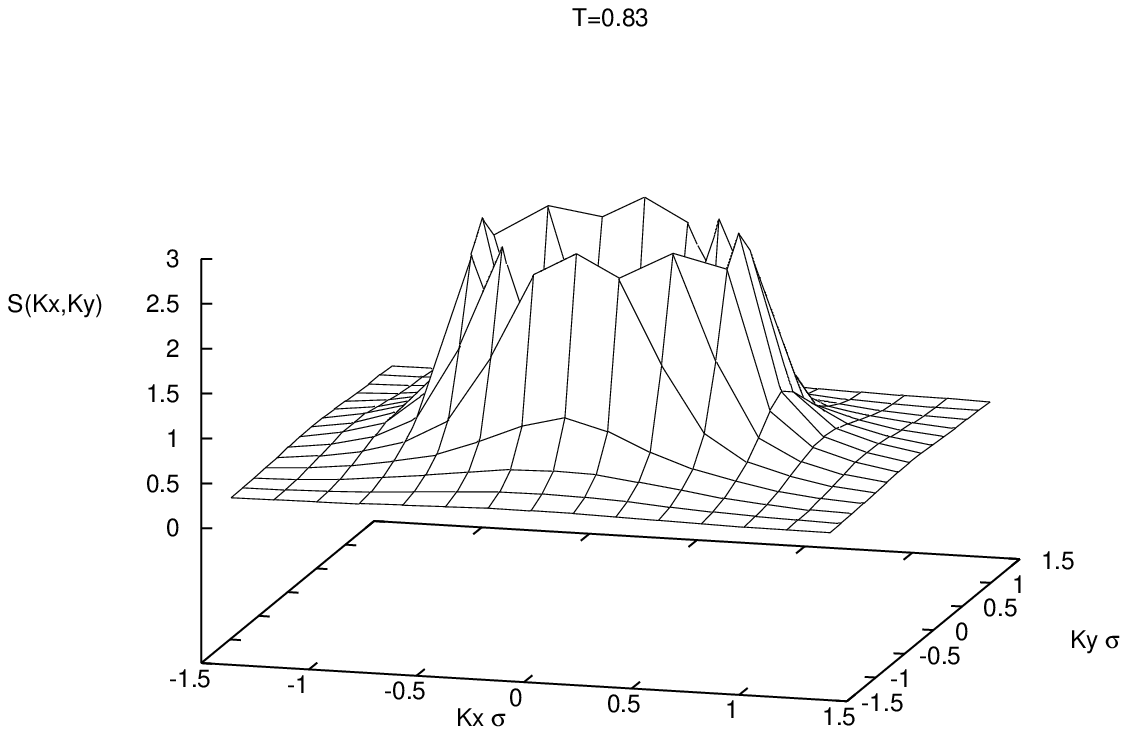}{8cm}{7cm}{0}
\end{center}
\vspace{-1cm}
\caption{\footnotesize{Structure factor at different temperatures for $\rho=0.4$ adopting the MC$_{cl}$ technique. Upper panels shows temperatures below the critical one, while the bottom panels are at temperatures higher than $T_c^s$.}}\label{Sgrid}
\end{figure}

Even if the interparticle interaction potential is characterized by a spherical symmetry, we see that, in the microphase region of the phase diagram, the pattern shape can show a breaking of symmetry. Such feature is particularly evident in the stripe case where the particle domains are aligned along a fixed direction, even if such direction is selected at random. To make clear this aspect we have shown in figure \ref{Sgrid} the structure factor computed on a grid of $\protect{\bf{k}}$ vectors (the mesh of which depends on the simulation box sides as previously explained), using the MC$_{cl}$, at different temperatures. Below the critical temperature (upper panels) the pattern shape has a two-fold symmetry as we can argue from the presence of a modulation peak (and its specular one) only along a single direction. At $T=0.32$ secondary peaks can also be seen due to the harmonics. The height of the modulation peaks decreases with temperature, their width, instead, is limited by the finite size of the simulation cell. At the temperatures higher than the critical one (bottom panels), the structure factor shape changes: modulation peaks appear in different direction till to form a ring at small $k$. The presence of the ring means that, also at high temperatures, enhanced density fluctuations are present and their characteristic length scale is linked to the inverse of the radius of such ring according to the formula $\lambda\sim2\pi/k$. Some anisotropy exhibited by the height of the structure factor along the ring is still present at $T\gtrsim T_c$ and this can be interpreted as a measure of the persistence of such density fluctuations,
persistence which can be very long (data shown in figure \ref{Sgrid} are, in fact, obtained via simulation whose run length is $5\,10^6$ MC steps). Similar results are obtained with PT technique.

\section{Random phase approximation and simple model for phase behavior\label{Microphase: mean field models and phase behavior}}

\begin{figure}[h!]
\begin{center}
\mfig{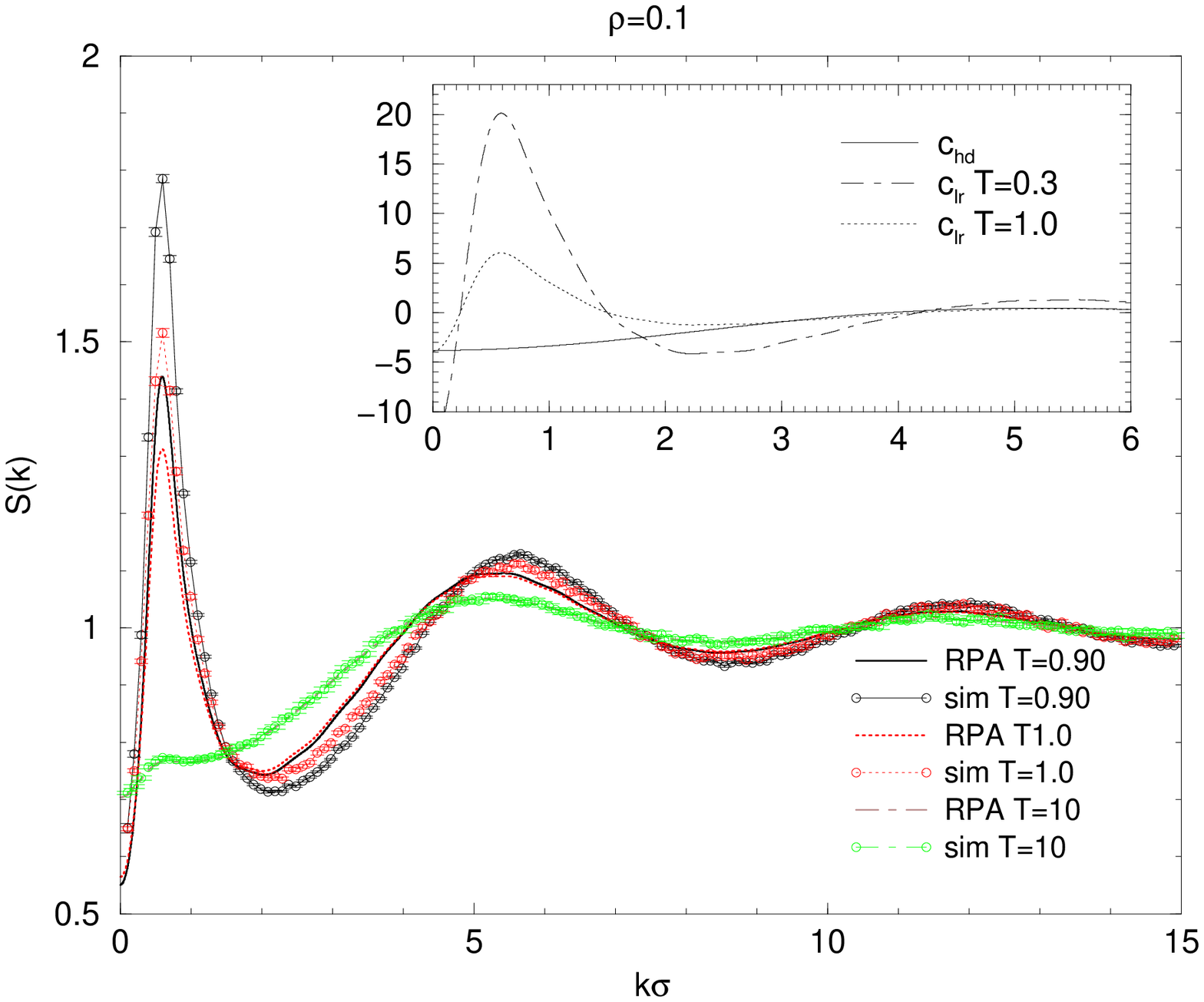}{8cm}{8cm}{0}\mfig{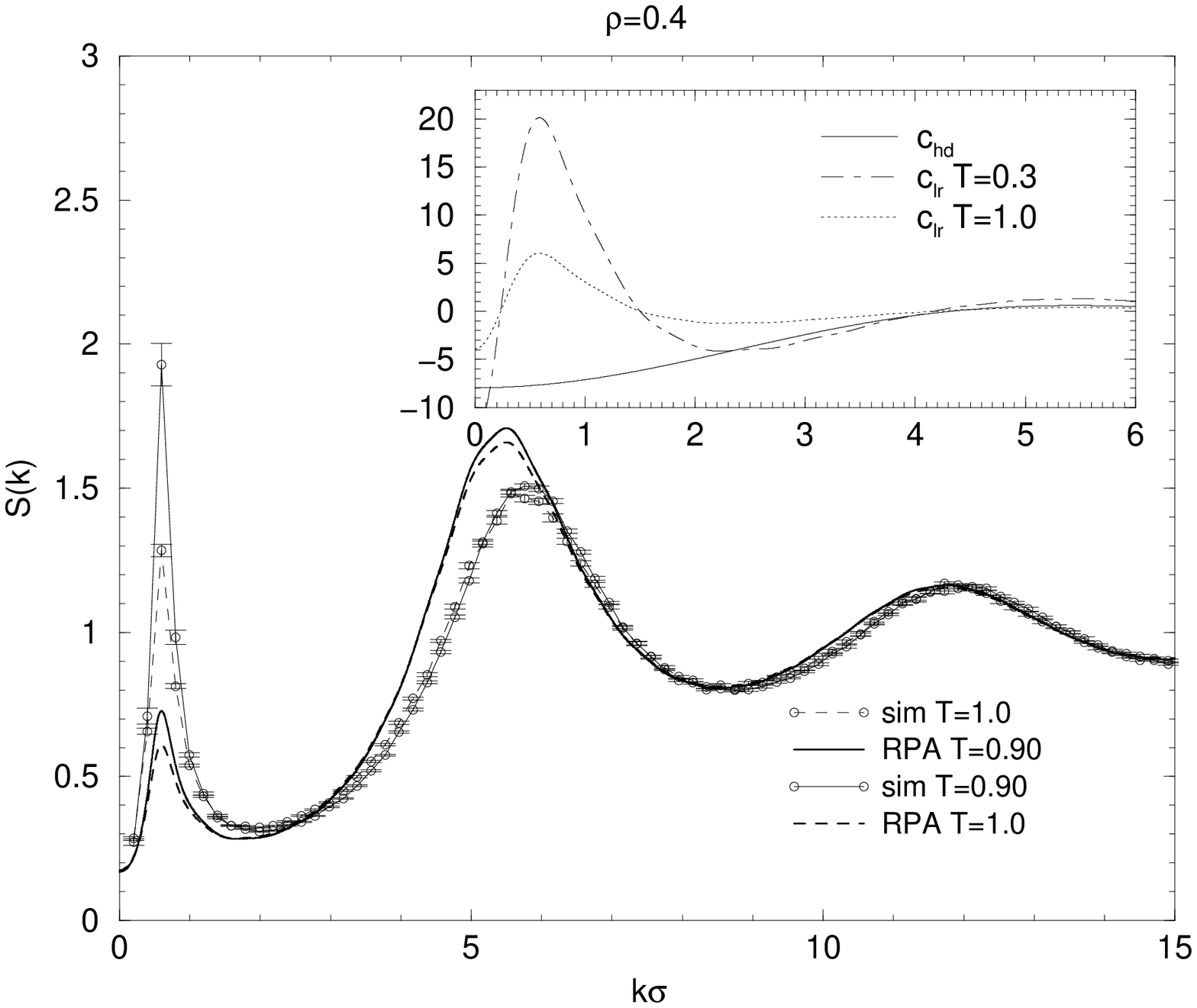}{8cm}{8cm}{0}
\end{center}
\vspace{-1cm}
\caption{\footnotesize{RPA predictions about the structure factor compared with simulation data for different densities and temperatures in the homogeneous regime; in the insets we show a comparison between the hard disk direct correlation function and the contribution due to long-range term of the potential.}}\label{RPA}
\end{figure}

\begin{figure}[h!]
\begin{center}
\mfig{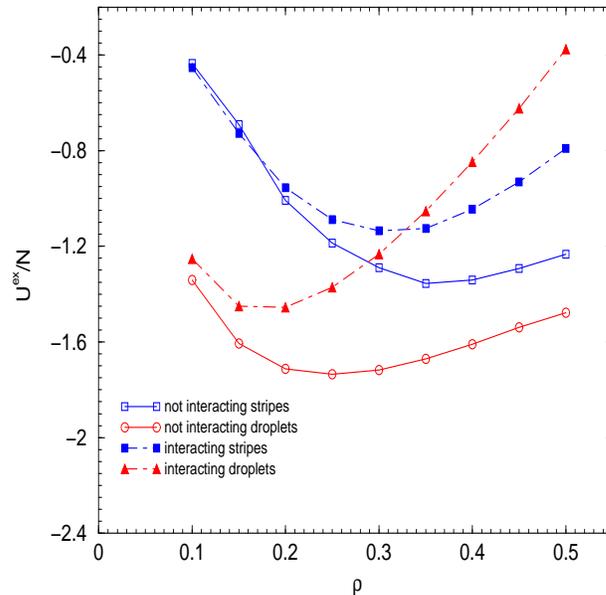}{8cm}{8cm}{0}
\end{center}
\vspace{-1cm}
\caption{\footnotesize{Model predictions about the droplets-stripes transition, with and without cluster-cluster repulsive interaction.}}\label{MF1}
\end{figure}

The static structure factor $S(k)$ gives many important details on the structure of a fluid; $S(k)$ defined as:
\begin{equation}
S(k)=1+\rho h(k),
\end{equation}
where $h(r)$ is the total correlation function, can be also expressed, via the Ornstein-Zernicke equation as 
\begin{equation}
S(k)=\frac{1}{1-\rho c(k)}\label{OZ},
\end{equation}
in which the direct correlation function $c(k)$ appears. Equation \ref{OZ} is more useful because, for potentials with steep repulsion plus a weak tail, we can use the following decomposition:
\begin{equation}
c(k)=c_{sr}(k)+c_{lr}(k)\label{OZ1},
\end{equation}
making a distinction between a short-range contribution ($c_{sr}$) due to the hard disk potential and a long-range one due to $U_{lr}$. For $c_{sr}$ we adopted the expression given in \cite{baus86}, which comes from simulation data fitting. For the long-range contribution we have used the random phase approximation (RPA) $c_{lr}\simeq -U_{lr}(k)/k_BT$. In figure \ref{RPA} we compare this approximate structure factor for few thermodynamic states at high temperature with simulation data. The theoretical as well as the simulation $S(k)$ exhibit a large peak for $k\sigma<1$ which is connected at high temperature to a characteristic length scale of fluctuations which at low temperature becomes the characteristic wave vector of the patterns. The insets show a comparison between $c_{hd}$ and $c_{lr}$: for $k\sigma<1$ $c_{lr}$ has a much stronger dependence on $k$ than $c_{hd}$, so that the structure is mainly dominated by the long-range potential (see also \cite{sear98}). Simulations however show stronger density fluctuations compared with the RPA results. The agreement between simulation and RPA about the modulation peak is better at higher temperature, since the approximation of a weak potential tail is better fulfilled.
Within the RPA, the locus of points where $S(k)$ diverges is the spinodal that we reported in figure \ref{Tc}, where we also reported the critical temperature estimated for $\rho=0.1$ and $\rho=0.4$ from the specific heat. The spinodal curve is quite similar to that of a classical fluid which undergoes only a fluid-fluid separation and it exhibits a critical density around $\rho\sim0.3$. From our preliminary simulation we found a transition from circular domains to stripe domains just around $\rho\sim(0.3\div0.35)$, but we have not yet explored in detail this transition region.
\begin{figure}[h!]
\begin{center}
\mfig{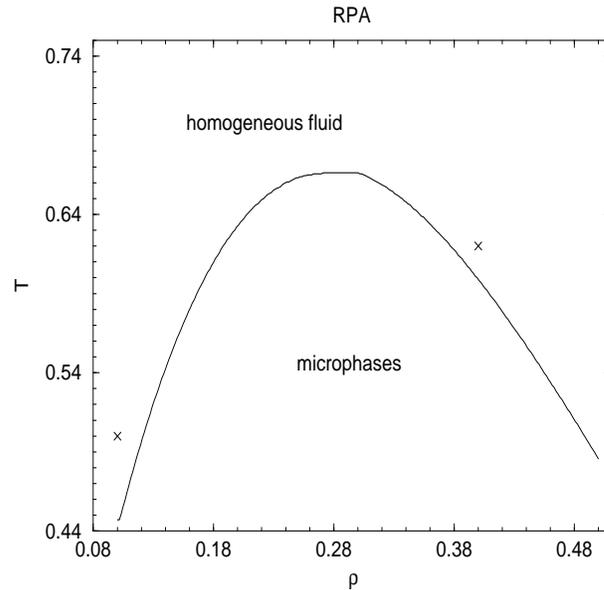}{8cm}{8cm}{0}\label{Tc}
\end{center}
\vspace{-1cm}
\caption{\footnotesize{Spinodal curve obtained studying the divergence of the structure factor according to equations \ref{OZ} and \ref{OZ1}. Crosses are simulation results obtained from the study of specific heat (see previous section).}}\label{Tc}
\end{figure}
At low $T$ due to competition between long-range repulsion and short-range attraction a fluid-fluid separation is forbidden, but particles arrange themselves on smaller clusters to decrease reciprocal repulsion. Increasing density, however, the cluster-cluster repulsive interaction is not negligible anymore and above a certain density a new pattern becomes favoured in order to minimize repulsion, leading to a morphology change that in our case is a passage from droplets to stripes. Let us assume that the transition is dominated by energetic effects and we introduce a very simplified model in order to predict the stability of the two patterns. We treat separately the interactions among particles which belong to the same cluster and the interactions among particles belonging to different clusters, in order to enlighten the role of reciprocal interaction among clusters. The first kind of interaction is referred to with the subscript letter $c$, while the latter ones with the subscript letters $cc$. The excess internal energy per particle can be so expressed as:
\begin{equation}
U^{ex}/N= u_c+ \chi\cdot u_{cc}\label{origin},
\end{equation}
where $\chi=N_{cl}z/2$ ($N_{cl}$ number of clusters and $z$ coordination number of the clusters, i.e. 6 for droplets supposed on a triangular lattice and 2 for stripes supposed on a regularly spaced grid) is proportional to the number of first neighbours clusters. Our goal is to compute such quantity in the density range $\rho (0.1\div0.5)$ for each pattern (droplets and stripes).\\
Once chosen the pattern we suppose the clusters are identical, that is all the droplets have the same diameter and all the stripes have the same width. We also suppose that the average distance $d$ among clusters is fixed and set equal to the value obtained by RPA. Since $N$ (total number of particles), $A$ simulation box area and $d$ average distance among clusters are fixed, we can obtain, for each pattern, the number of clusters $N_{cl}$, the number of particle $n$ and the density $\rho_{cl}$ inside each cluster and their area $S_{cl}$.\\
An approximate expression for $u_c$ and $u_{cc}$ are:
\begin{equation}
u_c=(U^{ex}/n)_c=\frac{1}{2} \,\rho_{cl}\int U(r)g(r)\rmd\protect{\bf{r}}\label{Uc}
\end{equation}
\begin{equation}
u_{cc}=(U^{ex}/n)_{cc}=\frac{\rho_{cl}}{S_{cl}}\int_{S_{cl}^1}\int_{S_{cl}^2} U(r)g_n^{(2)}(\protect{\bf{r}}^1,\protect{\bf{r}}^2)\rmd\protect{\bf{r}}^1\rmd\protect{\bf{r}}^2\label{Ucc}
\end{equation}
where $U(r)$ is the interaction potential adopted of equation \ref{kac}, the integral is over the cluster domain and the superscript numbers 1 e 2 underline the fact that we refer to two different clusters. The expressions for $u_c$ and $u_{cc}$ are obtained in analogy of the excess internal energy that can be computed for a homogeneous fluid subject to a pair-wise potential. To be very simple we neglect correlations, i.e. we put
\begin{equation}
\begin{array}{l}
g(r)\sim 1\\
g_n^{(2)}(\protect{\bf{r}}^1,\protect{\bf{r}}^2)\sim 1\\
\end{array}
\end{equation}
 Remembering that $U^{ex}=-\frac{\partial \ln Q}{\partial \beta}$ ($Q$ partition function) and that the free energy is $A=~-~\ln Q/\beta$, we can obtain an estimation of the free energy as:
\begin{equation}
A=\frac{1}{\beta}\int_{\beta_o}^{\beta}U^{ex}d\beta=\frac{U^{ex}\cdot(\beta-\beta_o)}{\beta},
\end{equation}
since the excess energy does not depend on temperature in our approximation. $\beta_o$ is a reference state with respect to which we compute the change of free energy. Assuming that the reference state is the ideal one, such approximation leads to $A^{ex}= U^{ex}$, so that what we will say for the excess internal energy can be immediately related to the free energy.\\
If in equation \ref{origin} we neglect the contribution $u_{cc}$ due to interaction among clusters we find that the circular domain should be preferred for all densities (see figure \ref{MF1}); on the contrary including $u_{cc}$ we see that the results are totally different: there exist different density regions in which droplet patterns and striped patterns appear respectively preferred (figure \ref{MF1}), i.e. there exist a ``critical density $\rho_o$'' such that for $\rho<\rho_o$ the system manages to minimize its excess internal energy (that, in our model, is equal to the free energy) setting particles on a droplet pattern, while for $\rho>\rho_o$ it achieves such energy minimization setting particles on a striped pattern. Inside this  very simplified model the passage between droplets and stripes happens around $\rho_o\sim0.33$, consistent with our simulation evidence.

\section{Microphases induced by external modulating potentials\label{Simulations with external modulating potential}}

\begin{figure}[h!]
\begin{center}
\mfig{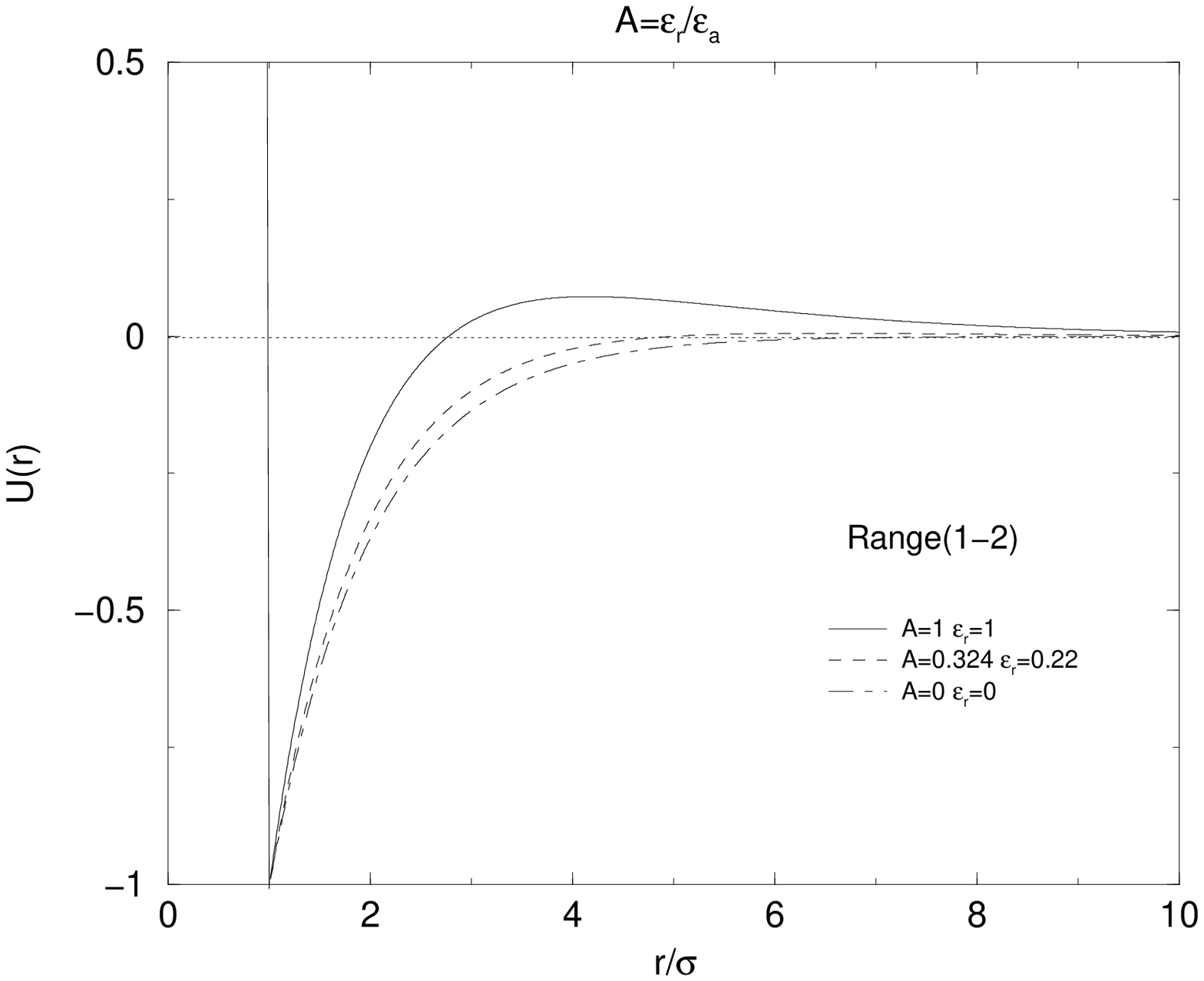}{8cm}{8cm}{0}\mfig{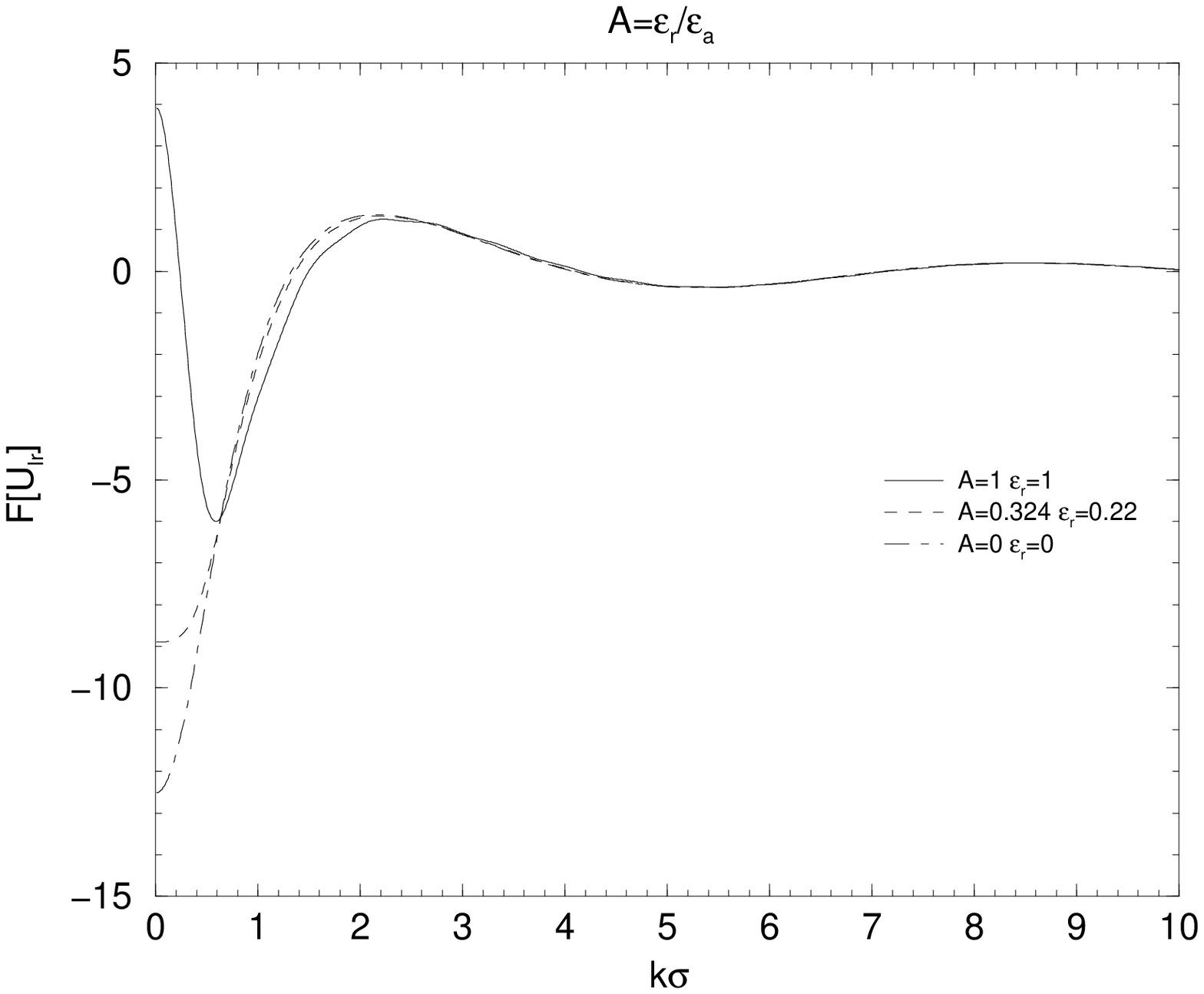}{8cm}{8cm}{0}
\end{center}
\vspace{-1cm}
\caption{\footnotesize{Interaction potential with different value of the parameters relative to the strength of the attractive and repulsive component (left) and its Fourier transform (right). The ranges of interactions are $R_a=1\sigma$ and $R_r=2\sigma$.}}\label{MF}
\end{figure}

If the intensity $\epsilon_r$ of the repulsive long-range potential in equation \ref{Ulong} is small the system displays a standard liquid-vapor phase transition with a critical temperature which is a decreasing function of $\epsilon_r$. The value of $\epsilon_r$ at which a fluid subject to competing interactions stops to undergo a standard liquid-vapor transition favouring a liquid-modulated phase is called Lifshitz point (LP) \cite{pini00}. In RPA this point corresponds to a change of convexity of the Fourier transform of the interaction potential at zero wave vector. Theoretical studies of $3D$ systems \cite{selk} emphasize that, approaching LP (but without entering the microphase region yet), the inverse of the compressibility of the fluid has a very small value over an extended region in density and in temperature. That means that the system manages to achieve large density fluctuations without undergoing a phase transitions. Such a behavior suggests that just in this region the presence of an external modulation should greatly affect the system. In this section we discuss such situation in bidimensional systems. The interaction potential is the same as in equation \ref{kac}: the ranges $R_a$ and $R_r$ have not been changed, while the intensity of the short-range and long-range interactions are $\epsilon_a=0.679$ and $\epsilon_r=0.22$. Such parameters correspond to the LP. In figure \ref{MF} we show the interaction potential and its Fourier transform for different values of $A=\epsilon_r/\epsilon_a$.
We added, along the $x$ direction, an external modulating potential of the form :
\begin{equation}
U_m=-\epsilon_m\sum_i^N\cos^2(\frac{2\pi x_i}{\lambda})
\end{equation}
in which the strength $\epsilon_m$ is in units of the contact value $U_c$ (defined in section \ref{Microphases: model system and computational details}), $x_i$ refers to the $i$-particle abscissa; the period of the function is $P=\lambda/2$ that, in the present case, we have taken equal to half of the simulation box side. Simulations have been done adopting MC$_{cl}$ scheme (described in section \ref{Microphases: model system and computational details}) in a canonical ensemble and with $N=400$ particles.\\
Adopting the random phase approximation described in section \ref{Microphase: mean field models and phase behavior} we estimate the critical point position at $\rho_c\sim0.28$ and $T_c\sim0.95$. From simulation we have verified that the fluid, without modulating potential, undergoes a standard liquid-vapor transition below the critical point.
\begin{figure}[h!]
\begin{center}
\mfig{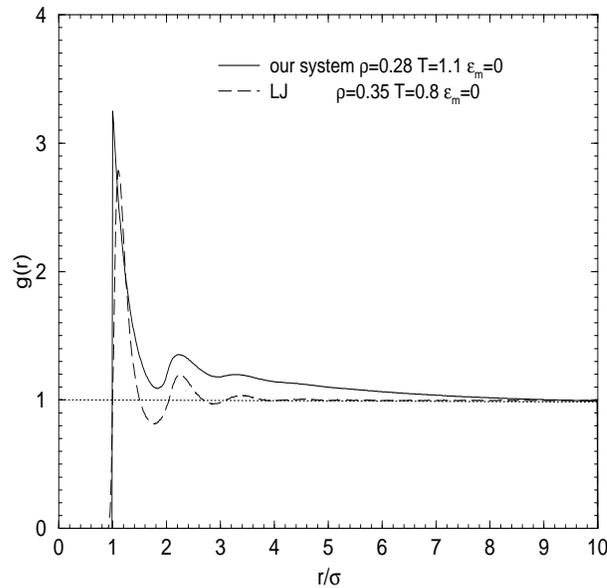}{8cm}{8cm}{0}
\end{center}
\vspace{-1cm}
\caption{\footnotesize{Radial distribution function for our system at the thermodynamic state $\rho=0.28$ $T=1.1$ next LP  compared with a typical behavior of the same quantity for a classical LJ fluid (without external modulation).}}\label{G_LP}
\end{figure}
\begin{figure}[h!]
\begin{center}
\mfig{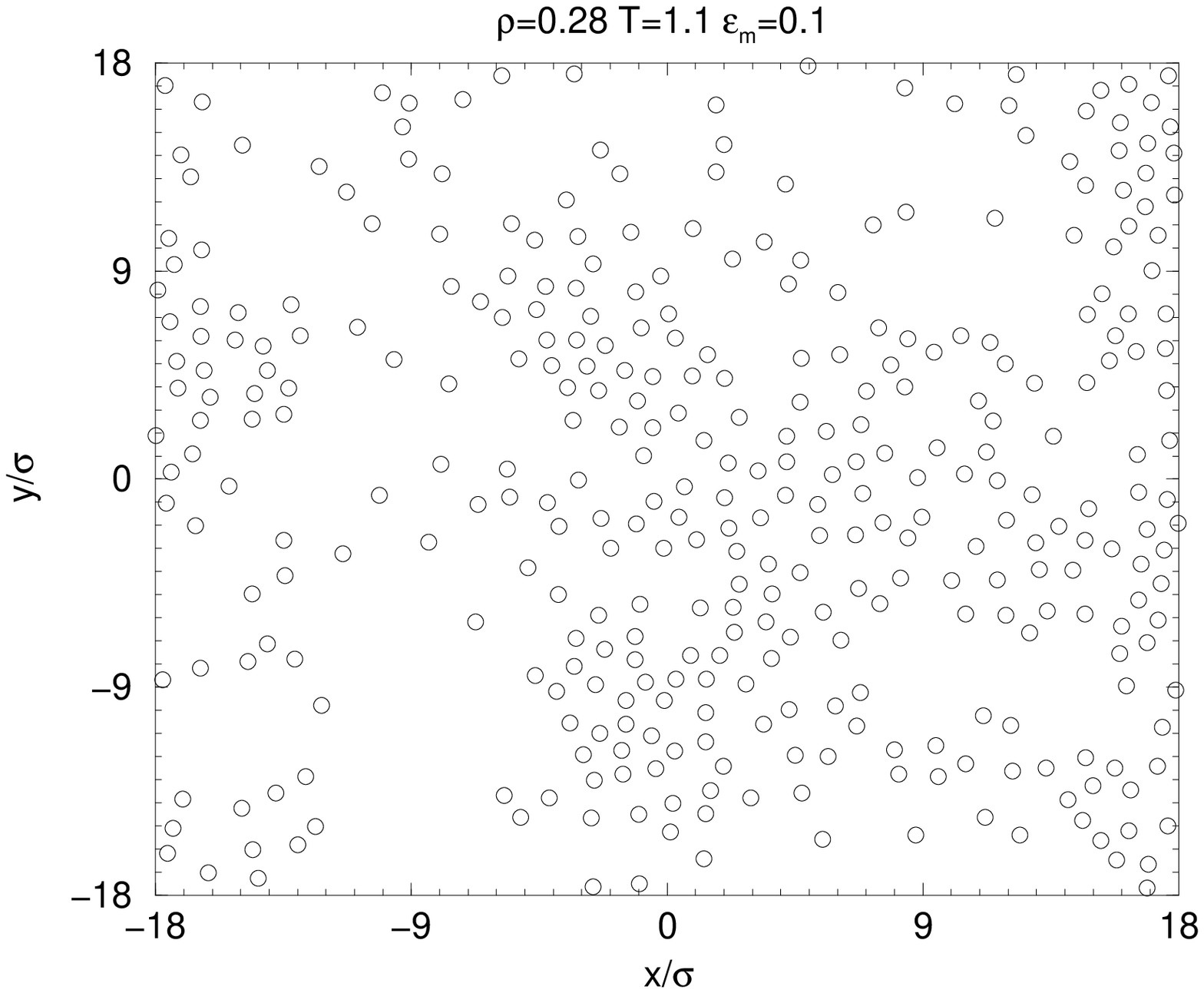}{8cm}{8cm}{0}\mfig{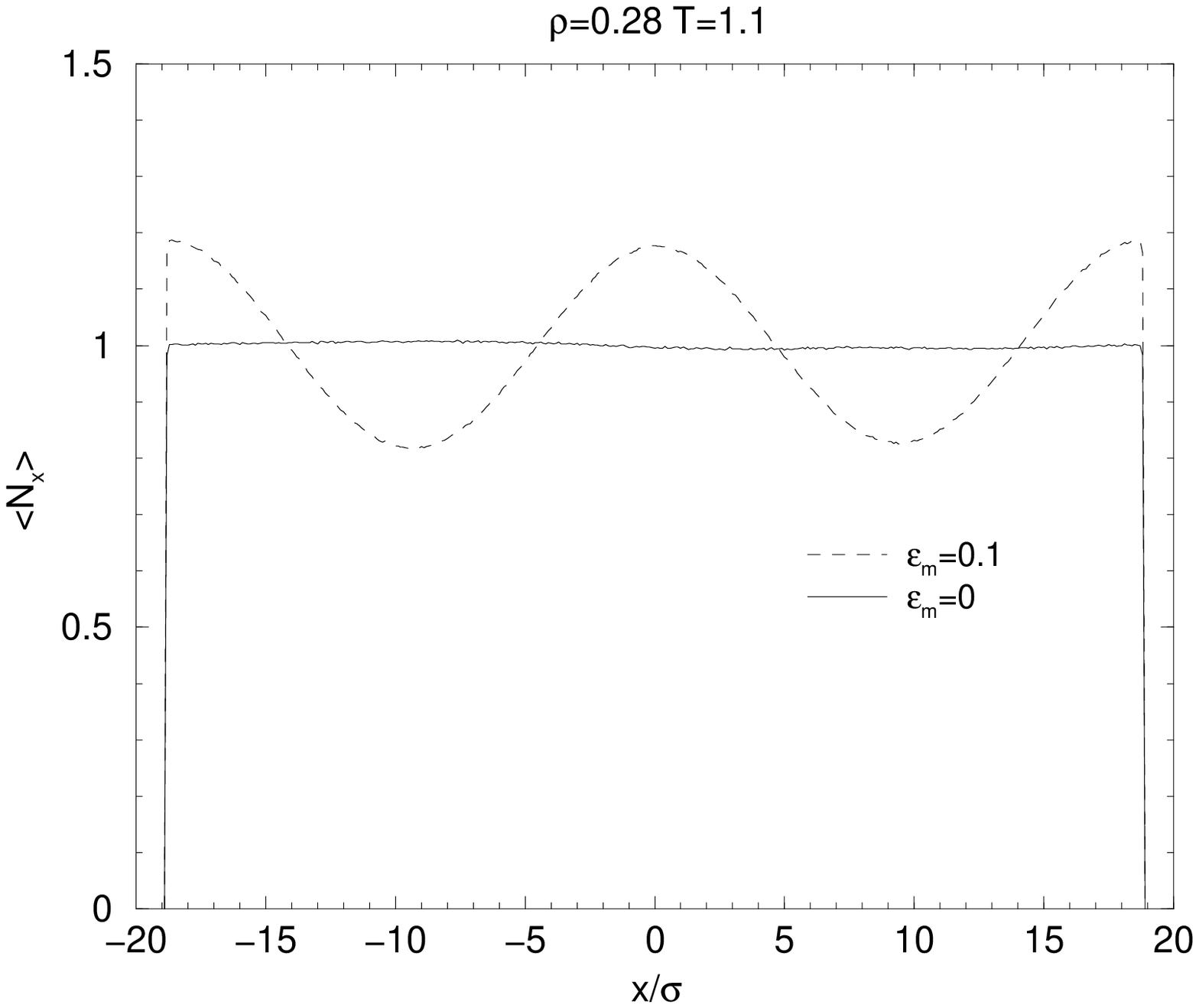}{8cm}{8cm}{0}
\end{center}
\vspace{-1cm}
\caption{\footnotesize{Left: snapshot relative to our system for $\rho=0.28$ $T=1.1$ and $\epsilon_m=0.1$. Right: number or particles versus position along the modulation axis normalized to the mean number of particles without external potential. }}\label{ss_LP}
\end{figure}

\begin{figure}[h!]
\begin{center}
\mfig{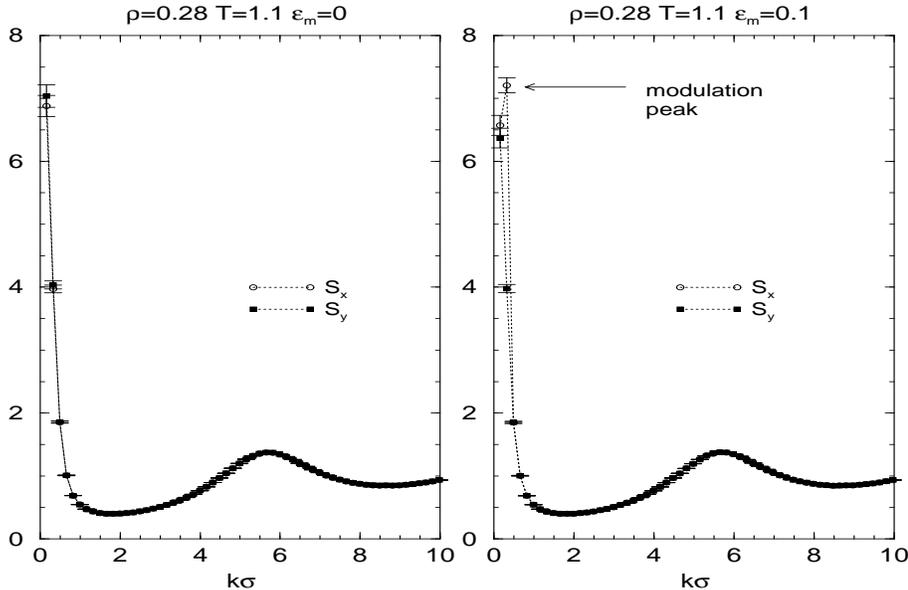}{12cm}{8cm}{0}
\end{center}
\vspace{-1cm}
\caption{\footnotesize{Static structure factor along $x$ and $y$ directions without external modulation (left panel) and with external modulation (right panel); the position of the modulation peak is indicated. Dotted lines are only a guide to eyes. }}\label{S_LP}
\end{figure}
\begin{figure}[ht!]
\begin{center}
\mfig{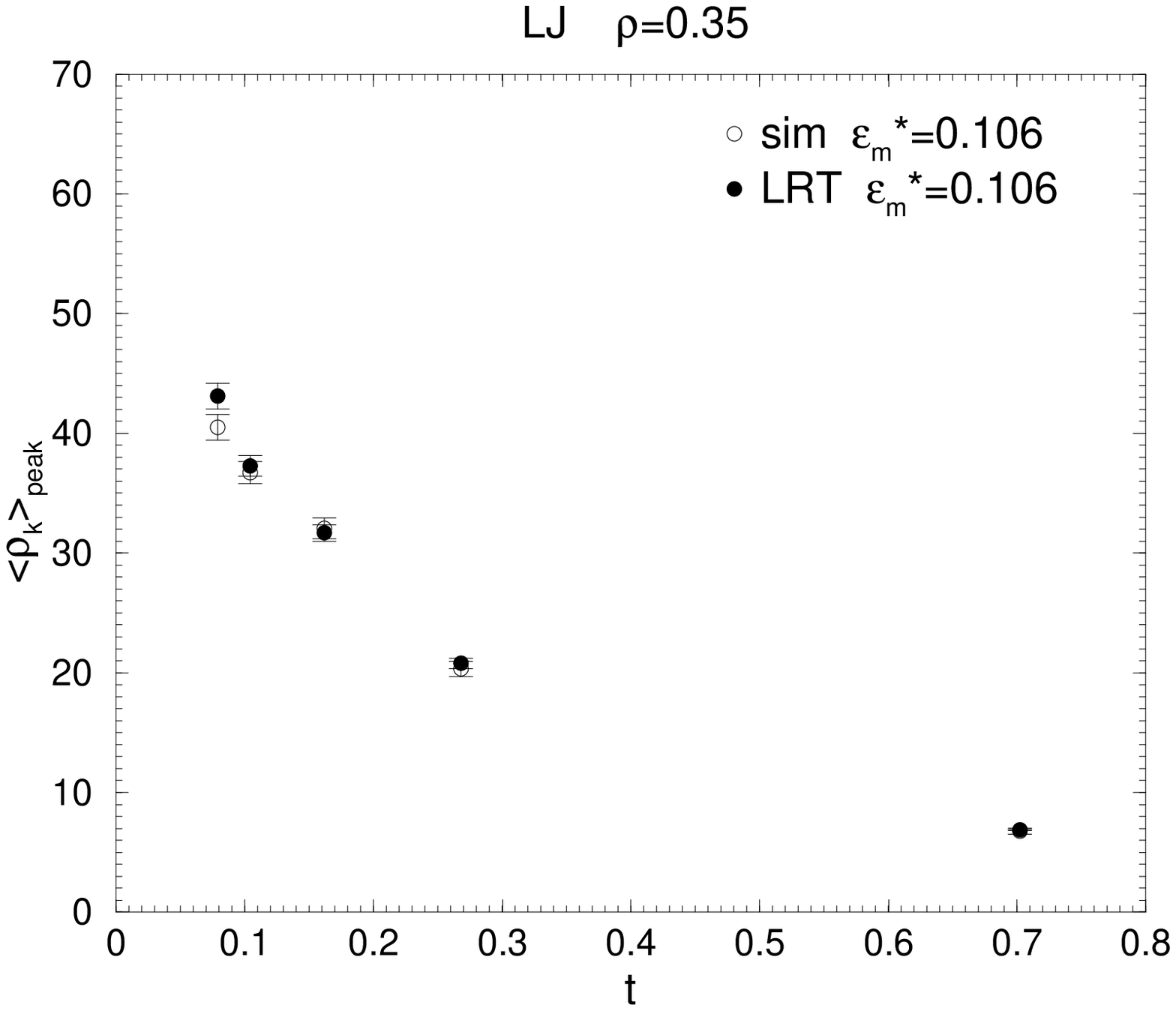}{8cm}{8cm}{0}\mfig{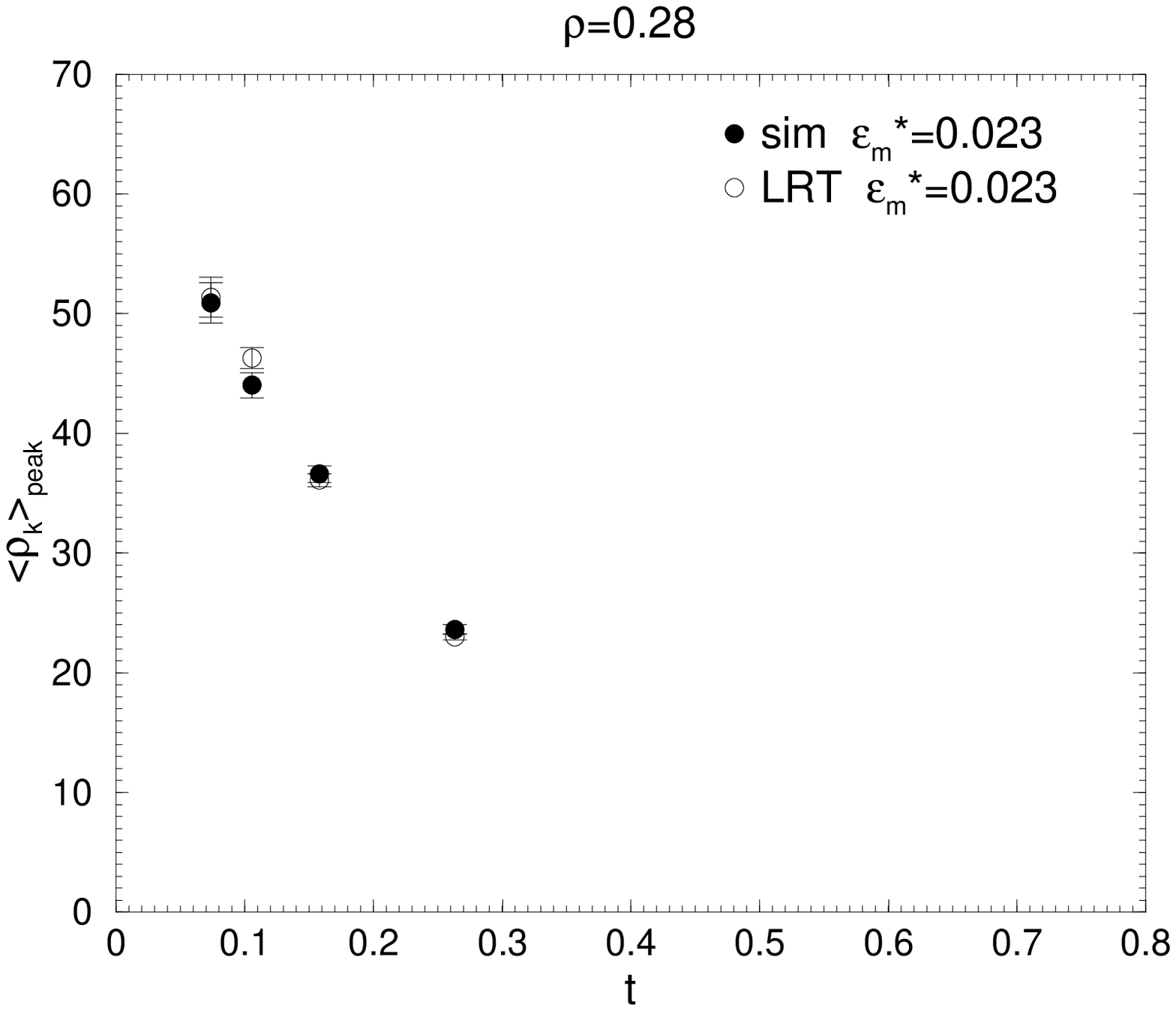}{8cm}{8cm}{0}
\mfig{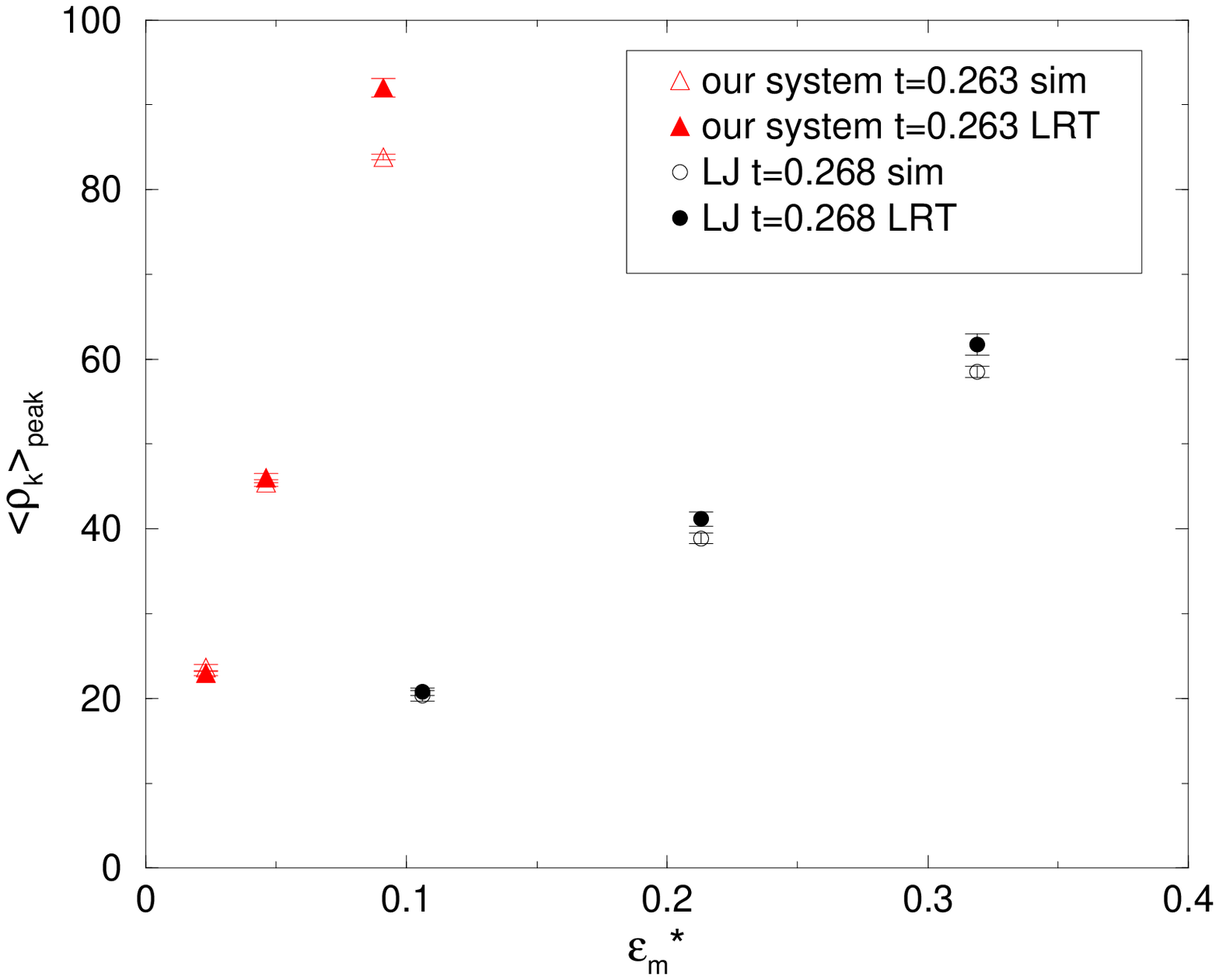}{8cm}{8cm}{0}
\end{center}
\vspace{-1cm}
\caption{\footnotesize{Upper panels: LRT predictions and simulation results about the Fourier transform of the density profile along the modulation axis in correspondence at the modulation wave vector for a LJ fluid (left) and for our system (right), versus reduced temperature $t=(T-T_c)/T_c$ at fixed $\epsilon_m^{\star}$. Bottom panel: LRT and simulation results versus $\epsilon_m^{\star}$ at fixed temperature. For each system temperatures and strengths of the modulating potential are expressed in units of their own critical temperature.}}\label{lrt}
\end{figure}
We studied a temperature range rather close to the estimated $T_c$ along the critical isochore. Without external modulation  and $T\gtrsim T_c$ the fluid phase is characterized by a radial distribution function very slowly decaying to unity, as we can see from figure \ref{G_LP} where we plotted $g(r)$ for the thermodynamic state $\rho=0.28$ $T=1.1$ in comparison with a case relative to a bidimensional Lennard-Jones system (LJ) confirming the great influence of the density fluctuations in this particular region.  Switching on the modulation, particles tend to organize in stripes located in correspondence of the minimum of the external potential (see figure \ref{ss_LP} for a snapshot of the system and a typical density profile along the modulation direction). Obviously a stripe pattern becomes better defined as $\epsilon_m$ increases. In figure \ref{S_LP} an example of structure factor is reported with and without external modulation. For $\epsilon_m=0$ $x$ and $y$ directions are equivalent since the fluid is homogeneous and isotropic; the large values of $S(k)$ at small $k$ are due to the proximity of the critical point. For $\epsilon_m=0.1$ at $k\sigma\simeq 0.33$ (corresponding to the modulation wave vector of the external potential) $S(k)$ has particularly increased with respect to the counterpart computed along the $y$ axis.\\
To test the effect of the density fluctuations close to the LP, we have compared our system with the bidimensional LJ, studying the response to an external modulation at different temperatures and for various intensities of the external potential. To simplify the comparisons we have also expressed temperatures and modulation strength in terms of the critical temperature for each system. LJ data are collected in table \ref{LJ_tab}, where $T\equiv k_BT/w$ and $\epsilon_m\equiv\epsilon_m/w$ ($w$ depth well), $t=(T-T_c^{LJ})/T_c^{LJ}$, $\epsilon_m^{\star}=\epsilon_m w/T_c^{LJ}$; estimation of the critical point for a LJ system in $2 D$ can be found in literature, according to which $\rho_c^{LJ}\sim 0.35$ and $T_c^{LJ}\sim0.47$ \cite{rovere93} (for a critical review about $2D$ LJ see also \cite{wilding92,wilding95}. Our system data are collected in table \ref{KAC_tab} where $t=(T-T_c)/T_c$ and  $\epsilon_m^{\star}=\epsilon_m U_c/T_c$.\\
From simulations we computed the Fourier transform of the density profile along the modulation direction ($<\rho_k>^{sim}$). Then we applied the linear response theory (LRT) according to which
\begin{equation}
<\rho_k>=<\rho_k>_o-\beta \rho S(k)_o U_m(k),\label{LRT}
\end{equation}
where the subscript $o$ denotes a canonical average over the unperturbed system and $U_m(k)$ is the Fourier component of the external potential at wave vector $k$; since the latter corresponds to a homogeneous fluid phase we can set $<\rho_k>_o=0$. The static structure factor which appears in the equation \ref{LRT} has been estimated via simulation at $\epsilon_m=0$ (having verified that $<\rho_k>_o^{sim}=0$ within the errors). In the upper panels of figure \ref{lrt} we plotted the results obtained for LJ and for our system relative to the modulation wave vector as a function of temperature. In both cases we see that, for small modulation strength $\epsilon_m$, LRT is a very good approximation even close to the critical point. In the bottom panel of figure \ref{lrt} we plotted a comparison between LRT predictions and simulation data as a function of $\epsilon_m^{\star}$ for a given $t$: the response to the external potential of the system with competing interaction is stronger than it is in the LJ system and the deviation from the linear theory starts at a smaller value of $\epsilon_m^{\star}$. For example in our case at $\epsilon_m^{\star}=0.091$\hspace{0.2cm} $<\rho_k>_{peak}\,\simeq83$ and LRT is already not very accurate, while in the LJ system at $\epsilon_m^{\star}=0.106$\hspace{0.2cm} $<\rho_k>_{peak}\,\simeq20$ and LRT still works very well. Obviously as $\epsilon_m^{\star}$ increases too much LRT is not more accurate also for the LJ system.
\begin{table}
\caption{\footnotesize{Temperature and external modulation strength, used in simulations for the LJ system, in units of the well depth and in units of the LJ critical temperature. See the text for explicit definitions.}}\label{LJ_tab}
\vspace{0.2cm}
\begin{indented}
\item[]\begin{tabular}{| p{1.2cm}| p{1.2cm}| p{1.2cm}| p{1.2cm}|}\hline
\hspace{0.5cm}$T$ & \hspace{0.5cm}$t$ & \hspace{0.5cm}$\epsilon_m$ &\hspace{0.5cm} $\epsilon_m^{\star}$\\\hline\hline
   0.507 &  0.079 & 0.05 &  0.106 \\
   0.519 &  0.104 & 0.10 &  0.213 \\
   0.546 &  0.162 & 0.15 &  0.319 \\ 
   0.596 &  0.268 &      &        \\ 
   0.800 &  0.702 &      &        \\\hline
\end{tabular}
\end{indented}
\end{table}
\begin{table}
\caption{\footnotesize{Temperature and external modulation strength, used in simulations using our competing interaction potential, in units of both the contact value $U_{c}$ and the critical temperature. See the text for explicit definitions.}}\label{KAC_tab}
\vspace{0.2cm}
\begin{indented}
\item[]\begin{tabular}{| p{1.2cm}| p{1.2cm}| p{1.2cm}| p{1.2cm}|}\hline
\hspace{0.5cm}$T$ & \hspace{0.5cm}$t$ & \hspace{0.5cm}$\epsilon_m$ &\hspace{0.5cm} $\epsilon_m^{\star}$\\\hline\hline
   1.020 &  0.074 &  0.1 &  0.023 \\
   1.050 &  0.105 &  0.2 &  0.046 \\
   1.100 &  0.158 &  0.4 &  0.091 \\
   1.200 &  0.263 &      &        \\ \hline
\end{tabular}
\end{indented}
\end{table}

\section{Conclusions\label{conclusions}} 

In this work we have presented detailed simulation data relative to the study of spontaneous pattern formation in bidimensional system with competing short-range attractive and long-range repulsive, showing that the PT technique is a promising approach, since this technique reaches equilibrium configurations faster than other algorithms and it improves the statistics quality. At the densities of our computations, we have observed the formation of droplet-like or stripe-like patterns. The location of the transition temperature from the modulated region to the homogeneous phase has been clearly identified through the position of the peak in the specific heat. \\
 We have also analysed how the arising of patterns affects structural quantities such as $g(r)$ and $S(k)$. The radial distribution function exhibits a long-range modulation due to the underlying pattern, which is doomed to disappear as temperature increases.
The main feature, into the static structure factor, is the appearance of peaks (modulation peaks) at small wave vector. In particular for $\rho=0.4$ at $T\leq T_c$ $S(\protect{\bf{k}})$, computed on the grid of small wave vectors, is anisotropic in $\protect{\bf{k}}$ and has a two-fold symmetry due to the breaking of symmetry that the system undergoes to, while at $T>T_c$ $S(\protect{\bf{k}})$ is isotropic with a unique strong peak at small wave vector. This signals that strong density fluctuations are present at temperature above $T_c$ where no patterns are present. Besides, in the striped case we observe in the $S(\protect{\bf{k}})$ profile, at low temperatures, the presence of many peaks at wave vectors which are multiple of the modulation period. The origin of such harmonics can be traced back to the rather abrupt density variation between its maximum and minimum values. This effect could be checked through scattering experiments. Some harmonics peaks are also present at $\rho=0.1$\ in the droplet phase but these peaks are weaker and wider.\\
In a small system, like ours, the direction of the striped pattern appears to be stabilized by the periodic boundary conditions; on the contrary in a very large system we expect that the direction of the stripes will fluctuate in time. It should be an important test determining the time scale of such fluctuations since they might influence properties like the birefrigence that could be experimentally measured.\\
For the homogeneous state the random phase approximation works quite well in giving the position of the fluctuation peak and it is semiquantitative for its amplitude. Even if RPA predicts only the existence of a region in which the fluid is unstable with respect to a density modulation, without giving any information on the pattern shape, it is interesting to note the presence of a critical point in the spinodal curve given by RPA and the critical density is into the region in which, via simulation, we observed the passage from droplets to stripes. At last a very simple model, to mimic the droplet-stripe transition, has been described. The most relevant aspect is that the shape of the pattern is an expedient to optimize and reduce cluster reciprocal repulsion. The development of a microscopic theory able to describe the modulated phases is an important open issue.\\
Finally we have studied the fluid behavior when it is close to its Lifshitz point but it is still in a homogeneous phase which it is strongly affected by density fluctuations. We have so analysed the system response to the presence of an external agent, such as a modulating potential. Comparing the results with those deriving from a standard LJ fluid, subject to the same external modulation, we have seen that our system is much more affected by the action of the field. In other words, next to the  Lifshitz point the fluid experiments large density fluctuations which can be easily driven into a stable microphase by the presence of a very weak external potential.

\ack
We thank D. Pini and D. Galli for the helpful discussions.\\
This work has been supported by the INFM Parallel Computing Project and by the INFM Research Project PAIS2001-Sez.G.

\section*{References}

\bibliography{biblio}

\end{document}